\documentclass[iop]{emulateapj}
\usepackage{epsfig,amsfonts,amsmath,graphicx,natbib,apjfonts,afterpage}
\def\ra#1#2#3{#1$^{\rm h}$ #2$^{\rm m}$ #3$^{\rm s}$}
\def\dec#1#2#3{$#1^\circ #2' #3''$}
\def\nod{\nodata}
\def\swift{{\it Swift}}

\def\chandra{{\it Chandra}}
\def\grb{GRB\,150101B}
\newcommand{\be}{\begin{equation}}
\newcommand{\ee}{\end{equation}}

\def\ein{1}
\def\az{2}
\def\nyu{3}
\def\nw{4}
\def\oh{5}
\def\har{6}
\def\hub{7}
\def\car{8}
\def\war{9}
\def\lei{10}
\def\nrao{11}
\def\ber{12}
\def\psu{13}
\def\cal{14}
\def\dar{15}
\def\gw{16}

\shorttitle{Short GRB 150101B}
\shortauthors{Fong~et.~al.}

\begin{document}

\title{The Afterglow and Early-Type Host Galaxy of the Short GRB\,150101B at $z=0.1343$}

\author{ 
W.~Fong\altaffilmark{\ein}$^,$\altaffilmark{\az},
R.~Margutti\altaffilmark{\nyu}$^,$\altaffilmark{\nw},
R.~Chornock\altaffilmark{\oh},
E.~Berger\altaffilmark{\har},
B.~J.~Shappee\altaffilmark{\hub}$^,$\altaffilmark{\car},
A.~J.~Levan\altaffilmark{\war},
N.~R.~Tanvir\altaffilmark{\lei},
N.~Smith\altaffilmark{\az},
P.~A.~Milne\altaffilmark{\az},
T.~Laskar\altaffilmark{\nrao}$^,$\altaffilmark{\ber},
D.~B.~Fox\altaffilmark{\psu},
R.~Lunnan\altaffilmark{\cal},
P.~K.~Blanchard\altaffilmark{\har},
J.~Hjorth\altaffilmark{\dar},
K.~Wiersema\altaffilmark{\lei}, \\
A.~J.~van der Horst\altaffilmark{\gw},
D.~Zaritsky\altaffilmark{\az}
}

\altaffiltext{\ein}{Einstein Fellow}
\altaffiltext{\az}{Steward Observatory, University of Arizona, 933 N. Cherry Avenue, Tucson, AZ 85721}
\altaffiltext{\nyu}{Center for Cosmology and Particle Physics, New York University, 4 Washington Place, New York, NY 10003, USA}
\altaffiltext{\nw}{Northwestern University, Department of Physics and Astronomy, 2145 Sheridan Road, Evanston, IL 60208, USA}
\altaffiltext{\oh}{Astrophysical Institute, Department of Physics and Astronomy, 251B Clippinger Lab, Ohio University, Athens, OH 45701, USA}
\altaffiltext{\har}{Harvard-Smithsonian Center for Astrophysics, 60 Garden Street, Cambridge, MA 02138, USA}
\altaffiltext{\hub}{Hubble Fellow}
\altaffiltext{\car}{Carnegie Observatories, 813 Santa Barbara Street, Pasadena, CA 91101, USA}
\altaffiltext{\war}{Department of Physics, University of Warwick, Gibbet Hill Road, Coventry CV4 7AL, UK}
\altaffiltext{\lei}{Department of Physics \& Astronomy, University of Leicester, University Road, Leicester LE1 7RH, UK}
\altaffiltext{\nrao}{National Radio Astronomy Observatory, 520 Edgemont Road, Charlottesville, VA 22903, USA}
\altaffiltext{\ber}{Department  of  Astronomy,  University  of  California,  501 Campbell Hall, Berkeley, CA 94720-3411, USA}
\altaffiltext{\psu}{Department of Astronomy and Astrophysics, Pennsylvania State University, 525 Davey Laboratory, University Park, PA 16802}
\altaffiltext{\cal}{Department of Astronomy, California Institute of Technology, 1200 East California Boulevard, Pasadena, CA 91125, USA}
\altaffiltext{\dar}{Dark Cosmology Centre, Niels Bohr Institute, University of Copenhagen, Juliane Maries Vej 30, DK-2100 Copenhagen Ø, Denmark}
\altaffiltext{\gw}{Department of Physics, 725 21st Street NW, The George Washington University, Washington, DC 20052}

\begin{abstract}

We present the discovery of the X-ray and optical afterglows of the short-duration \grb, pinpointing the event to an early-type host galaxy at $z=0.1343 \pm 0.0030$. This makes \grb\ the most nearby short GRB with an early-type host galaxy discovered to date. Fitting the spectral energy distribution of the host galaxy results in an inferred stellar mass of $\approx 7 \times 10^{10}~M_{\odot}$, stellar population age of $\approx 2-2.5$~Gyr, and star formation rate of $\lesssim 0.4$~$M_{\odot}$~yr$^{-1}$. The host of \grb\ is one of the largest and most luminous short GRB host galaxies, with a $B$-band luminosity of $\approx 4.3L^*$ and half-light radius of $\approx 8$~kpc. \grb\ is located at a projected distance of $7.35 \pm 0.07$~kpc from its host center, and lies on a faint region of its host rest-frame optical light. Its location, combined with the lack of associated supernova, is consistent with a NS-NS/NS-BH merger progenitor. From modeling the evolution of the broad-band afterglow, we calculate isotropic-equivalent gamma-ray and kinetic energies of $\approx 1.3 \times 10^{49}$~erg and $\approx (6-14) \times 10^{51}$~erg, respectively, a circumburst density of $\approx (0.8-4) \times 10^{-5}$~cm$^{-3}$, and a jet opening angle of $\gtrsim 9^{\circ}$. Using observations extending to $\approx 30$~days, we place upper limits of $\lesssim (2-4) \times 10^{41}$~erg~s$^{-1}$ on associated kilonova emission. We compare searches following previous short GRBs to existing kilonova models, and demonstrate the difficulty of performing effective kilonova searches from cosmological short GRBs using current ground-based facilities. We show that at the Advanced LIGO/VIRGO horizon distance of 200~Mpc, searches reaching depths of $\approx 23-24$~AB~mag are necessary to probe a meaningful range of kilonova models.

\end{abstract}

\section{Introduction}

The afterglows and host galaxies of short-duration gamma-ray bursts (GRBs; $T_{\rm 90}\lesssim 2$~s; \citealt{kmf+93}) provide critical information about their explosion properties and progenitors. Modeling of their afterglows from the radio to X-ray bands has led to a median isotropic-equivalent energy scale of $\approx\,10^{51}$~erg and low explosion environment densities of $\approx 10^{-3}-10^{-2}$~cm$^{-3}$ \citep{ber07,nak07,nkg+12,ber14,fbm+15}. These properties are in stark contrast to long-duration GRBs ($T_{\rm 90}\gtrsim 2$~s), which have a median isotropic-equivalent energy scale of $\approx 10^{53}$~erg \citep{fks+01,bkf03,fb05,gbb+08,nfp09,lbt+14} and substantially higher circumburst densities of $\approx 0.1-100$~cm$^{-3}$ \citep{pk02,yhs+03}. Furthermore, the association of long GRBs with Type Ic supernovae (e.g., \citealt{hb12}) and their exclusive association to star-forming galaxies \citep{dkb+98,ldm+03,fls+06,wbp07} has helped to establish their origin from massive stars.

\begin{figure*}
\centering
\includegraphics*[angle=0,width=\textwidth]{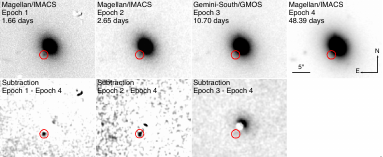}
\caption{Optical $r$-band observations of the afterglow of \grb. Imaging was obtained with IMACS mounted on the Magellan/Baade 6.5-m telescope and GMOS mounted on the Gemini-South 8-m telescope at four separate epochs, $\delta t$ = 1.66, 2.65, 10.70, and 48.39 days (observer frame). Digital image subtractions between the last epoch and each of the previous epochs reveal a fading optical afterglow at RA=\ra{12}{32}{05.094} and Dec=\dec{-10}{56}{03.00} (J2000) with a $1\sigma$ positional uncertainty of $0.24''$ (red circle; $5\sigma$ radius). Due to saturation of the host galaxy core and differences in seeing between the epochs, the position of the afterglow in the subtractions is contaminated by residuals from the host galaxy subtraction, necessitating PSF photometry to properly measure the afterglow brightness. Image subtractions have been smoothed with a 2-pixel Gaussian, and the scale and orientation of all panels are as shown in the fourth panel.
\label{fig:subpanel}}
\end{figure*}

In comparison, short GRBs lack associated supernovae \citep{ffp+05,hsg+05,hwf+05,sbk+06,ktr+10,ber14}, strongly suggestive of an older stellar progenitor. Furthermore, the detection of an $r$-process kilonova associated with the short GRB\,130603B was the first direct evidence linking them to neutron star-neutron star and/or neutron star-black hole merger progenitors (NS-NS/NS-BH; \citealt{bfc13,tlf+13}). The spatial distribution of short GRBs within their hosts, together with their weak correlation with local star formation and stellar mass \citep{fbf10,ber10,fb13,tlt+14}, are also fully consistent with expectations for NS-NS/NS-BH mergers \citep{fwh99,bsp99,bpb+06,zrd09,brf14}.

Critical to the study of short GRB progenitors are their host galaxy properties and demographics. While early studies based on a small number of events found a dominance of early-type host galaxies \citep{zr07}, a later study focused on $36$~short GRBs found that $\approx 20-40\%$ of events are hosted by early-type host galaxies, with a dominance of star-forming hosts \citep{fbc+13}. However, the star-forming hosts of short GRBs have lower specific star formation rates, higher luminosities, and higher metallicities than the star-forming hosts of long GRBs \citep{ber09}. Short GRBs have been discovered over the redshift range $z \approx 0.12-1.3$ (c.f., \citealt{ber14}) with the single outlier GRB\,090426A at $z=2.609$ \citep{aap+09,lbb+10}. Only two short GRBs\footnote{The long-duration GRBs\,060505 and 060614 are associated with star-forming galaxies at $z=0.0894$ and $z=0.125$, respectively. These events lack associated supernovae to deep limits, suggesting that they do not have massive star progenitors \citep{ocg+07,gfp+06}.} have been discovered at $z \lesssim 0.2$: GRB\,050709 ($z=0.161$) and GRB\,080905A ($z=0.1218$), both of which have star-forming host galaxies \citep{ffp+05,hwf+05,rwl+10}. A third event, GRB\,061201, is potentially associated with a star-forming galaxy at $z=0.111$; however, this association is less robust \citep{sdp+07,fbf10}. In general, events at low redshift offer an opportunity to study their explosion and host galaxy environments in great detail.

Here, we report the discovery of the optical and X-ray afterglows of the short \grb, which localizes the event to an early-type host galaxy at $z=0.1343 \pm 0.003$ ($D=636.7$~Mpc). This makes \grb\ the most nearby short GRB discovered in an early-type host galaxy to date. An earlier work concentrated on broad-band modeling of the host galaxy to uncover an active galactic nucleus (AGN) at the host center \citep{xfw+16}. Here, we seek to provide a comprehensive view of this event. In Section~2, we present the discovery of the optical and X-ray afterglows, as well as optical and near-IR host galaxy  observations. In Section~3, we analyze the broad-band afterglow to infer the burst explosion properties, and place deep limits on supernova and kilonova emission. In Section~4, we determine the host galaxy redshift, stellar population and morphological properties, and the location of \grb\ within its host. In Section~5, we compare the properties of \grb\ and its host to previous short GRBs. In particular, we discuss the implications for the progenitors and for future kilonova searches. We draw conclusions in Section~6.

Unless otherwise noted, all magnitudes in this paper are in the AB system and are corrected for Galactic extinction in the direction of the burst using $E(B-V)=0.036$~mag and $R_V=3.1$ \citep{sfd98,sf11}. Reported uncertainties correspond to $68\%$ confidence, unless otherwise indicated. We employ a standard $\Lambda$CDM cosmology with $\Omega_M=0.286$, $\Omega_\Lambda=0.714$, and $H_0=69.6$ km s$^{-1}$ Mpc$^{-1}$.

\section{Observations of \grb}

\grb\ was detected on 2015 Jan 1 at 15:23 UT \citep{gcn17267} by the Burst Alert Telescope (BAT; \citealt{bbc+05}) on-board the \swift\ satellite, and by the Gamma-ray Burst Monitor (GBM) on-board the {\it Fermi} satellite at 15:23:35 UT \citep{gcn17267}. \swift/BAT localized the burst to RA=\ra{12}{32}{10.4} and Dec=$-$\dec{10}{58}{48} (J2000) with $2.4'$-radius precision ($90\%$ containment; \citealt{lsb+16}). The $15-150$~keV $\gamma$-ray emission consists of a single pulse with a duration of $T_{90}=0.012 \pm 0.009$~s \citep{lsb+16}, classifying \grb\ as a short-duration GRB with no extended emission. Employing a power law fit for the BAT spectrum gives a fluence of $f_{\gamma}=(6.1 \pm 2.2) \times 10^{-8}$~erg~cm$^{-2}$ and a photon index of $\Gamma_{\gamma}=1.4 \pm 0.6$ in the $15-150$~keV energy band ($\chi^2_{\nu}=0.7$; 90\% confidence; \citealt{lsb+16}). The fluence measured by the {\it Fermi}/GBM, which covers a wider energy range than \swift\ of $10-1000$~keV, is $f_{\gamma}=(1.09 \pm 0.14) \times 10^{-7}$~erg~cm$^{-2}$. The spectrum is best-fit by a power law with index $\Gamma_{\gamma}=1.70 \pm 0.09$ \citep{gcn17276}, consistent with the value measured from the \swift/BAT data. 

\subsection{Afterglow Observations and Discovery}
\label{sec:oadisc}

\tabletypesize{\footnotesize}
\begin{deluxetable*}{lccccccccc}
\tablecolumns{11}
\centering
\tablewidth{0pc}
\tablecaption{GRB\,150101B Optical and Near-IR Afterglow and Host Galaxy Observations
\label{tab:obsag}}
\tablehead {
\colhead {Date}			&
\colhead {$\delta t$}		&
\colhead {Telescope}		&
\colhead {Instrument}	&
\colhead {Band}		&
\colhead {Exposures}	&
\colhead {Afterglow} 		&
\colhead {$F_{\nu,{\rm opt}}$}		&
\colhead {Host}			&
\colhead {$A_{\lambda}$}	\\
\colhead {(UT)} 			&
\colhead {(days)}		&
\colhead {}			&
\colhead {}			&
\colhead {}			&
\colhead {(sec)}			&
\colhead {(AB mag)} 		&
\colhead {($\mu$Jy)}		&
\colhead {(AB mag)}		&
\colhead {(AB mag)}		          	           
}
\startdata
2015 Jan 3.297 & $1.66$ & Magellan/Baade 	& IMACS	& $r$	& $8 \times 150$ 	& $23.01 \pm 0.17$		& $2.27 \pm 0.36$ &   \nod           & 0.094 \\
2015 Jan 4.289	& $2.65$ & Magellan/Baade 	& IMACS	& $r$	& $10 \times 120$	& $23.53 \pm 0.26$		& $1.41 \pm 0.34$ & \nod          & 0.094 \\
2015 Jan 4.298  & $2.66$ & VLT			 & HAWK-I & $J$ 	& $12 \times 60$ &	$>22.3$				& $<4.37$		 	&  \nod	    & 0.029 \\
2015 Jan 4.310  & $2.67$ & VLT 			& HAWK-I & $H$ 	& $12 \times 60$ &   $>21.4$				& $<10.0$			     & \nod	    & 0.018 \\
2015 Jan 4.322  & $2.68$ & VLT 			& HAWK-I & $K$	& $12 \times 60$ & 	$>21.5$				& $<9.12$			& \nod 	   & 0.012 \\
2015 Jan 4.359 & $2.72$ & VLT & 			HAWK-I & $Y$		& $22 \times 60$ & 	$>20.5$				& $<22.9$			& \nod		& 0.043 \\
2015 Jan 11.261 & $9.62$ & TNG$^{a}$ 		& NICS 	& $J$	& 				& $>21.7$				& $<7.59$ 		     &  \nod	    & 0.029 \\
2015 Jan 12.341 & $10.70$ & Gemini-South 	& GMOS	& $r$ 	& $19 \times 90$ 	& $>24.2$ 			& $<0.76$			     & \nod	    & 0.094 \\
2015 Jan 16.201 & $14.56$ & TNG$^{a}$		& NICS	& $J$ 	& 				& $>22.4$				& $<3.98$ 		     &  \nod	    & 0.029 \\
2015 Jan 16.307 & $14.67$ & VLT 			& HAWK-I & $H$ 	& $22 \times 60$ 	& $>23.4$				& $<1.58$			     & \nod	    & 0.018 \\
2015 Jan 16.331 & $14.69$ & VLT 			& HAWK-I & $Y$ 	& $22 \times 60$ 	& $>23.6$				& $<1.32$				& \nod & 0.043 \\				
2015 Jan 21.533 & $19.89$    & UKIRT		& WFCAM & $J$	& $12 \times 200$     & $>22.4$			& $<3.98$ 		     & $15.47 \pm 0.05$  & 0.029 \\
2015 Jan 21.569 & $19.93$    & UKIRT		& WFCAM & $K$	& $12 \times 200$     & $>22.2$			& $<4.79$			     & $15.11 \pm 0.05$ & 0.012 \\
2015 Jan 30.508 & $28.87$    & UKIRT		& WFCAM & $J$	& $12 \times 200$	& $>23.0$				& $<2.29$			     & $15.55 \pm 0.05$ & 0.029 \\
2015 Jan 30.544 & $28.90$    & UKIRT		& WFCAM & $K$	& $12 \times 200$	& $>22.4$				& $<3.98$ 		     & $15.20 \pm 0.06$ &  0.012 \\
2015 Feb 11.320 & $40.70$ & HST 			& WFC3/IR & F160W & $3 \times 250$ 	& $>25.3$				& $<0.28$			     &  $15.09 \pm 0.01$ &  0.021 \\
2015 Feb 11.385 & $40.74$ & HST                    & WFC3/UVIS & F606W & $4 \times 350$ & $>25.1$                    	& $<0.33$			     & $16.57 \pm 0.01$ &   0.102 \\
2015 Feb 14.274 & $43.63$ & VLT 			& HAWK-I & $H$ 	& $22 \times 60$ 	& $>23.9$				& $<1.00$				& \nod		& 0.018 \\
2015 Feb 19.036 & $48.39$ & Magellan/Baade & IMACS	& $r$	& $8 \times 150$	& $>24.6$				& $<0.52$			     & \nod	     & 0.094 \\
2015 Feb 19.048 & $48.41$ & Magellan/Baade & IMACS	& $r$	& $3 \times 45$		& 					&              					    & $16.51 \pm 0.04$ & 0.094 \\
2015 Feb 19.053 & $48.41$ & Magellan/Baade & IMACS	& $g$	& $3 \times 60$		& 					&              					    & $17.42 \pm 0.04$ & 0.136 \\
2015 Feb 19.058 & $48.42$ & Magellan/Baade & IMACS	& $i$		& $3 \times 45$		&					&              					    & $16.08 \pm 0.04$ & 0.070 \\
2015 Feb 19.063 & $48.42$ & Magellan/Baade & IMACS	& $z$	& $3 \times 60$		&					&               				    & $15.77 \pm 0.05$ & 0.052
\enddata
\tablecomments{All upper limits correspond to $3\sigma$ confidence and uncertainties correspond to $1\sigma$ confidence. All values are corrected for Galactic extinction in the direction of the burst, $A_{\lambda}$ \citep{sf11}. \\
$^{a}$ From \citet{gcn17326}.
}
\end{deluxetable*}

\subsubsection{Optical Afterglow Discovery}

We obtained $r$-band observations of \grb\ with the Inamori Magellan Areal Camera and Spectrograph (IMACS) mounted on the Magellan/Baade 6.5-m telescope at a mid-time of 2015 Jan 03.297 UT ($\delta t=1.66$~days, where $\delta t$ is the time after the BAT trigger in the observer frame). The observations cover the entire BAT position (Figure~\ref{fig:subpanel} and Table~\ref{tab:obsag}). To assess variability of any optical sources within the BAT position, we obtained a second set of IMACS observations at $\delta t \approx 2.65$~days. Performing digital image subtraction between the two sets of observations using the ISIS software package \citep{ala00}, we find a faint point-like residual $\approx 3''$ southeast of a bright galaxy (Figure~\ref{fig:subpanel}), indicative of a fading optical source.

To assess the nature of this source and its association to \grb, we obtained two deeper sets of $r$-band observations with the Gemini Multi-Object Spectrograph (GMOS) mounted on the Gemini-South 8-m telescope at a mid-time of $\delta t \approx 10.70$~days, and with Magellan/IMACS at $\delta t \approx 48.39$~days (Table~\ref{tab:obsag}). We perform digital image subtraction between each of the initial epochs and the last set of Magellan observations at $\delta t \approx 48.39$~days, which serves as the deepest template. In the first two subtractions, a point-like residual source is present at the same position, and we consider this source to be the optical afterglow of \grb\ (Figure~\ref{fig:subpanel}).

To determine the position of the optical afterglow, we perform absolute astrometry between the earliest Magellan observations and the 2MASS catalog using the IRAF astrometry routines {\tt ccmap} and {\tt ccsetwcs}. Using 19 common point sources, the resulting $1\sigma$ astrometric tie uncertainty is $\sigma_{\rm 2MASS \rightarrow GRB} = 0.24''$. Since the position of the afterglow is contaminated by galaxy light in our discovery images, we use {\tt SExtractor}\footnote{http://www.astromatic.net/software/sextractor} to determine the afterglow position and uncertainty ($\sigma_{\rm GRB}$) in the subtracted images. The subtraction between the first and last epochs ($\delta t = 1.66$ and $48.39$~days, respectively) yields a positional uncertainty of $\sigma_{\rm GRB}=0.029''$. We therefore calculate an absolute optical afterglow position of RA=\ra{12}{32}{05.094}, Dec=\dec{-10}{56}{03.00} (J2000) with a total $1\sigma$ positional uncertainty of $0.24''$.

To measure the brightness of the afterglow in our first two Magellan observations, we perform PSF photometry using standard PSF-fitting routines in the IRAF/{\tt daophot} package. We note that the position of the afterglow is contaminated either by flux from the nearby galaxy in the observations, or by residuals from the saturated galaxy core in the subtracted images (Figure~\ref{fig:subpanel}), preventing accurate aperture photometry. We create a model PSF using $3-5$ bright, unsaturated stars in each field out to a radius of $4.5\theta_{\rm FWHM}$ from the center of each star, and then use this PSF to subtract $20$ additional stars in this field. The clean subtraction of these stars indicates an accurate model of the PSF for the observation. We then use the model PSF to subtract a point source fixed at the location of the afterglow. We calibrate all of the photometry relative to the Gemini observations at $\delta t=10.70$~days, which were taken in photometric conditions, using the most recent tabulated zeropoint. Taking into account the uncertainty in the zeropoint, as well as the Galactic extinction in the direction of the burst \citep{sf11}, we measure an afterglow brightness of $r_{\rm AB}=23.01 \pm 0.17$~mag and $r_{\rm AB}=23.53 \pm 0.26$~mag at $\delta t=1.66$ and $2.65$~days, respectively. Our optical observations and afterglow photometry, supplemented by published values from the literature, are summarized in Table~\ref{tab:obsag}.

To obtain upper limits on any transient optical flux from our late-time Gemini and Magellan observations, we inject fake sources of varying brightness to each observation at the position of the optical afterglow using IRAF/{\tt addstar}. We then perform PSF photometry and measure the uncertainty for each of the fake sources to determine the $3\sigma$ upper limit on the afterglow in each image. The resulting limits are listed in Table~\ref{tab:obsag}.

\subsubsection{Near-IR Upper Limits}

We obtained several observations of \grb\ with the High Acuity Wide field K-band Imager I (HAWK-I) mounted on the Very Large Telescope (VLT) in Chile in the $YJHK$ filters, spanning $\delta t \approx 2.66-43.63$~days (Table~\ref{tab:obsag}). We used standard routines as part of the {\tt esorex} pipeline for dark subtraction, flat-field correction, and stacking. For the series of $YJHK$ observations on 2015 Jan 4, guiding issues during these observations resulted in an elongated point spread function with a bright core, making the comparison with subsequent observations challenging, as well as reducing the resulting depth, with varying impact on each filter. In Table~\ref{tab:obsag}, we give point source limits from small aperture photometry. We obtained photometric calibration via 2MASS, although $Y$-band observations were calibrated against tabulated zero points and confirmed by a comparison of USNO-B1 I-band and 2MASS $J$-band observations of common sources. The elongated PSF means the reported limits are shallower than might be expected. We obtained additional $H$-band observations on 2015 Jan 16 and 2015 Feb 14. These provide deep limits on the presence of any transient emission, reaching $H_{\rm AB}\gtrsim 23.4$. The details of our observations, as well as $3\sigma$ limits, are listed in Table~\ref{tab:obsag}.

\tabletypesize{\small}
\begin{deluxetable*}{lccccc}
\centering
\tablecolumns{6}
\tablewidth{0pc}
\tablecaption{GRB\,150101B \chandra\ Afterglow Observations
\label{tab:xrayobs}}
\tablehead {
\colhead {$\delta t$}          &
\colhead {Exposure Time}           &
\colhead {$F_X$(0.3-10~keV)$^{a}$}                    &
\colhead {$F_{\nu}$(1~keV)}      &
\colhead {$\Gamma_X$} &
\colhead {N$_{\rm H,int}$} \\
\colhead {(days)}                 &
\colhead {(s)}                    &
\colhead {(erg cm$^{-2}$ s$^{-1}$)}        &
\colhead {($\mu$Jy)}   &
\colhead {}      &     
\colhead {(cm$^{-2}$)}                    
}
\startdata
$7.94$ & 14869 & $(1.16 \pm 0.12) \times 10^{-13}$ & $(1.08 \pm 0.11) \times 10^{-2}$ & $1.67 \pm 0.17$ & $\lesssim3.0 \times 10^{21}$ \\
$39.68$ & 14862 & $(2.08 \pm 0.50) \times 10^{-14}$ & $(1.94 \pm 0.47) \times 10^{-3}$ & $1.13 \pm 0.49$ & $\lesssim2.1 \times 10^{21}$
\enddata
\tablecomments{Uncertainties correspond to $1\sigma$ and upper limits correspond to $3\sigma$ confidence.\\
$^{a}$ Unabsorbed flux. }
\end{deluxetable*}

\subsubsection{X-ray Afterglow Discovery}
\label{sec:xray}

Observations with the X-ray Telescope (XRT; \citealt{bhn+05}) on-board \swift\ commenced at $\delta t=139.2$~ks and revealed an X-ray source with a UVOT-enhanced position of RA=\ra{12}{32}{04.96} and Dec=$-$\dec{10}{56}{00.7} (J2000), and an uncertainty of $1.8''$ in radius ($90\%$ containment; \citealt{gcn17268,ebp+07,ebp+09}). However, this X-ray source was found to remain constant in flux over the duration of the XRT observations, $\delta t \approx 139.2-2612$~ks. Furthermore, the position of this X-ray source is coincident with the nucleus of the host galaxy (see Section~\ref{sec:host}), which is a known galaxy 2MASX~J12320498-1056010 as catalogued in the NASA/IPAC Extragalactic Database \citep{hms+91}. No other fading source from the \swift/XRT observations was reported.

We analyze X-ray observations obtained at a mid-time of 2015 Jan 9.565 UT ($\delta t \approx 7.94$~days) with the Advanced CCD Imaging Spectrometer (ACIS-S) on-board the \chandra\ X-ray Observatory (ID: 16508492; PI: E.~Troja), with a net exposure time of $14.9$~ks. We retrieve the pre-processed Level 2 data from the \chandra\ archive and use the CIAO/{\tt wavdetect} routine to perform a blind search of X-ray sources at or near the optical afterglow position. We find an X-ray source located at RA=\ra{12}{32}{05.099} and Dec=$-$\dec{10}{56}{02.80} with a $1\sigma$ positional uncertainty of $0.38''$, coincident with the position of the optical afterglow (Section~\ref{sec:oadisc}). Within a $1.2''$-radius aperture, we compute a net count-rate of $9.5 \times 10^{-3}$~counts~s$^{-1}$ ($0.5-8$~keV) which corresponds to a significance of $\sim52\sigma$.

To assess the variability of this X-ray source, we analyze a second set of \chandra/ACIS-S observations obtained at a mid-time of 2015 Feb 10.305 UT ($\delta t \approx 39.68$~d; ID: 16708496, PI: A.~Levan). Fixing the position and aperture to those derived from the first \chandra\ epoch, we compute a net count-rate of $1.3 \times 10^{-3}$~counts~s$^{-1}$ ($0.5-8$~keV) which corresponds to a significance of $\sim6\sigma$. This indicates that the source has faded by a factor of $\approx 7$ between $\delta t \approx 7.94$ and $39.68$~days. Due to the fading of this source, as well as the spatial coincidence with the optical afterglow, we confirm this source as the X-ray afterglow of \grb. We note that the X-ray afterglow position and count rates are fully consistent with an independent analysis of the \chandra\ data by \citet{xfw+16}. 

To determine the flux calibration, we extract a spectrum from each of the \chandra\ epochs using CIAO/{\tt specextract}. We fit each of the data sets with an absorbed power law model characterized by photon index, $\Gamma_{\rm X}$, and intrinsic neutral hydrogen absorption column, $N_{\rm{H,int}}$, in excess of the Galactic column density in the direction of the burst, $N_{\rm H,MW}=3.48\times 10^{20}\,\rm{cm^{-2}}$; (typical uncertainty of $\sim 10\%$; \citealt{kbh+05,wlb11}), using Cash statistics. For the first epoch, we find a best-fitting spectrum characterized by $\Gamma_{\rm X}=1.67 \pm 0.17$ and a $3\sigma$ upper limit of $N_{\rm{H,int}}\lesssim 3.0 \times 10^{21}$~cm$^{-2}$ (C-stat$_\nu=0.85$ for $92$ d.o.f.). For the second epoch, the best-fitting spectrum is characterized by $\Gamma_{\rm X}=1.13 \pm 0.49$ and $N_{\rm{H,int}}\lesssim 2.1 \times 10^{21}$~cm$^{-2}$ (C-stat$_\nu=1.21$ for $13$ d.o.f.). We use the spectral parameters to calculate the count-rate-to-flux conversion factors, and hence the unabsorbed fluxes, in the $0.3-10$~keV energy band. The details of the \chandra\ observations, 0.3-10~keV fluxes, and best-fit spectral parameters are listed in Table~\ref{tab:xrayobs}. To enable comparison of the X-ray light curves to the optical and radio data, we also convert the X-ray fluxes to flux densities, $F_{\nu,X}$, at a fiducial energy of 1~keV ($F_{\nu,X} \propto \nu^{\beta_X}$ where $\beta_X \equiv 1-\Gamma_{\rm X}$); these values are listed in Table~\ref{tab:xrayobs}.

\subsubsection{Radio Upper Limit}

We observed the position of \grb\ with the Karl G. Jansky Very Large Array (VLA; Program 14A-344, PI: E.~Berger) starting at $\delta t \approx 5.73$~days at a mean frequency of $9.8$ GHz (upper and lower side-bands centered at $8.6$~GHz and $11.0$~GHz) using 3C286 and J1239-1023 for bandpass/flux and gain calibration, respectively. We follow standard procedures in the Astronomical Image Processing System (AIPS; \citealt{gre03}) for data calibration and analysis. The positions of the X-ray and optical afterglows are contaminated by radio emission from the host galaxy (see Section~\ref{sec:host}). To place a limit on any radio emission at the location of the afterglow, we use AIPS/{\tt IMSTAT} with the position fixed to that of the optical afterglow and calculate the average flux within a region fixed to the size of the radio beam. In this manner, we calculate a $3\sigma$ upper limit of $\lesssim1.5$~mJy for the radio afterglow of \grb.

\subsection{Host Galaxy Observations}
\label{sec:host}

\subsubsection{Optical Imaging and Identification}
\label{sec:optphot}

The optical afterglow of \grb\ is situated $\approx 3''$ southeast of a galaxy 2MASX~J12320498-1056010 as catalogued in the NASA/IPAC Extragalactic Database \citep{hms+91}. We obtained a set of late-time observations with Magellan/IMACS at a mid-time of $\delta t \approx 48.4$~days in the $griz$-bands (Table~\ref{tab:obsag} and Figure~\ref{fig:subpanel}). To measure the absolute position of the galaxy, we use the IRAF routines {\tt xregister} and {\tt wcscopy} to align the late-time observations with the afterglow discovery image, which is tied to 2MASS. This gives an absolute position of RA=\ra{12}{32}{04.973} and Dec=\dec{-10}{56}{00.50} (J2000) with a total $1\sigma$ positional uncertainty of $0.24''$, where the dominant source of uncertainty is the astrometric tie error. Taking into account the afterglow and host positional uncertainties, we calculate a relative offset between the location of the afterglow and galaxy center of $\delta$RA$=-1.89''$ and $\delta$Dec$=2.50''$, for a total angular offset of $\delta R = 3.07 \pm 0.03''$.

We use IRAF/{\tt apphot} to perform aperture photometry of the galaxy, where the zeropoint is calibrated to a standard star field taken on the same night at similar airmass. We use a $9''$-radius source aperture for all bands, where the aperture size is chosen to encompass the entire galaxy and maximize the signal-to-noise ratio. We utilize source-free regions surrounding the galaxy as background regions. The optical and near-IR photometry is listed in Table~\ref{tab:obsag}. We note that the photometric errors are dominated by uncertainties in the zeropoints.  

Given the angular separation, $\delta R$, and the optical galaxy brightness of $r_{\rm AB}=16.51 \pm 0.04$~mag, we calculate a probability of chance coincidence (c.f., \citealt{bkd02,ber10}) of $P_{\rm {cc}} \approx 4.8 \times 10^{-4}$. The low probability of chance coincidence, combined with the lack of any galaxies in the field with similarly low values, makes 2MASX~J12320498-1056010 the definitive host galaxy of \grb.

\subsubsection{Near-IR}

We obtained two sets of near-IR observations in the $JK$-bands with the Wide-field Camera (WFCAM; \citealt{caa+07}) mounted on the 4-m United Kingdom Infrared Telescope (UKIRT) starting at $\delta t\approx 19.89$ and $28.87$~days, respectively (Table~\ref{tab:obsag}). We obtained pre-processed images from the WFCAM Science Archive \citep{hcc+08} which are corrected for bias, flat-field, and dark current by the Cambridge Astronomical Survey Unit\footnote{http://casu.ast.cam.ac.uk/}. For each epoch and filter, we co-add the images and perform astrometry relative to 2MASS using a combination of tasks in Starlink\footnote{http://starlink.eao.hawaii.edu/starlink} and IRAF. To assess the presence of transient near-IR emission, we perform digital image subtraction between the two sets of observations in each filter and find no evidence for variability. Calibrated to 2MASS, we calculate $3\sigma$ limits of $J_{\rm AB}\gtrsim 22.4$~mag and $K_{\rm AB}\gtrsim 22.2$~mag at $\delta t \approx 19.9$~d (Table~\ref{tab:obsag}). Also listed are limits from near-IR observations published in the literature. We perform host galaxy photometry in the same manner as in the optical band (Section~\ref{sec:optphot}), and measure $J_{\rm AB}=15.47 \pm 0.05$~mag and $K_{\rm AB}=15.11 \pm 0.05$~mag (Table~\ref{tab:obsag}).

\subsubsection{HST Observations}
\label{sec:hstobs}

We retrieved pre-processed images from the {\it Hubble Space Telescope} ({\it HST}) archive\footnote{http://archive.stsci.edu/hst/} taken with the Wide Field Camera~3 (WFC3) at $\delta t \approx 40.7$~days (PI: N.~Tanvir) in two filters, F606W and F160W, which utilize the UV and IR channels of WFC3, respectively. We apply distortion corrections and combine the individual exposures using the {\tt astrodrizzle} package in PyRAF \citep{gon12}. For the WFC3/IR images, we use the recommended values of {\tt pixfrac}\,=\,1.0 and {\tt pixscale}\,=\,$0.13''$ pixel$^{-1}$, while for the WFC3/UVIS images, we use {\tt pixscale}\,=\,$0.04''$ pixel$^{-1}$. We use standard routines in IRAF to perform aperture photometry of the host galaxy and use PSF photometry to calculate upper limits; the results are listed in Table~\ref{tab:obsag}.

We perform relative astrometry between the {\it HST} and IMACS afterglow discovery images using IRAF astrometry routines. Taking into account the positional uncertainties of the afterglow and host, as well as the astrometric tie error between the {\it HST} and afterglow images, we calculate relative astrometric uncertainties of $\sigma_{\rm F160W}=0.07''$ and $\sigma_{\rm F606W}=0.04''$.

\subsubsection{Spectroscopy}

We obtained spectroscopic observations of the host on 2015 January 4.26 UT using Magellan/IMACS at a mean airmass of 2.0.  We obtained a pair of 600~sec exposures with the 300$+$17.5 grism covering $3900-10000$\,\AA\ at a spectral resolution of $\sim3.9$\,\AA.  The $0.7''$ slit was aligned at a position angle of 323$^\circ$ to go through both the nucleus of the host galaxy and the GRB position.  IMACS has an atmospheric dispersion compensator, so the relative flux scale is reliable.  We used standard tasks in IRAF to reduce the spectra and performed wavelength calibration using He-Ne-Ar arc lamps. We used observations of the standard stars LTT~1020 and LTT~3864 \citep{hsh+94} and custom IDL programs to apply a flux calibration and remove the telluric absorption bands. We calibrate the overall flux using the measured F606W, $g$-, $r-$, and $i$-band host photometry (Table~\ref{tab:obsag}).

\subsubsection{Archival Observations: a Low-luminosity AGN}
\label{sec:agn}

There are archival observations of the host of \grb, spanning the far-ultraviolet to the radio bands from GALEX, XMM, 2MASS, WISE, and the VLA. Using these observations, as well as \chandra\ and WSRT observations taken after the GRB, \citet{xfw+16} found that the broad-band spectral energy distribution matches that of a low-luminosity AGN. They classify the host galaxy as an X-ray bright, optically normal galaxy, with an X-ray luminosity of $\approx 7.5 \times 10^{42}$~erg~s$^{-1}$ ($0.2-10$~keV; \citealt{xfw+16}). We discuss further evidence for AGN activity in the optical spectrum in Section~\ref{sec:ssp}.

\section{Afterglow Properties}
\label{sec:sed}

\subsection{Energy Scale and Circumburst Density}

\begin{figure*}
\begin{minipage}[c]{\textwidth}
\tabcolsep0.0in
\includegraphics*[width=0.5\textwidth,clip=]{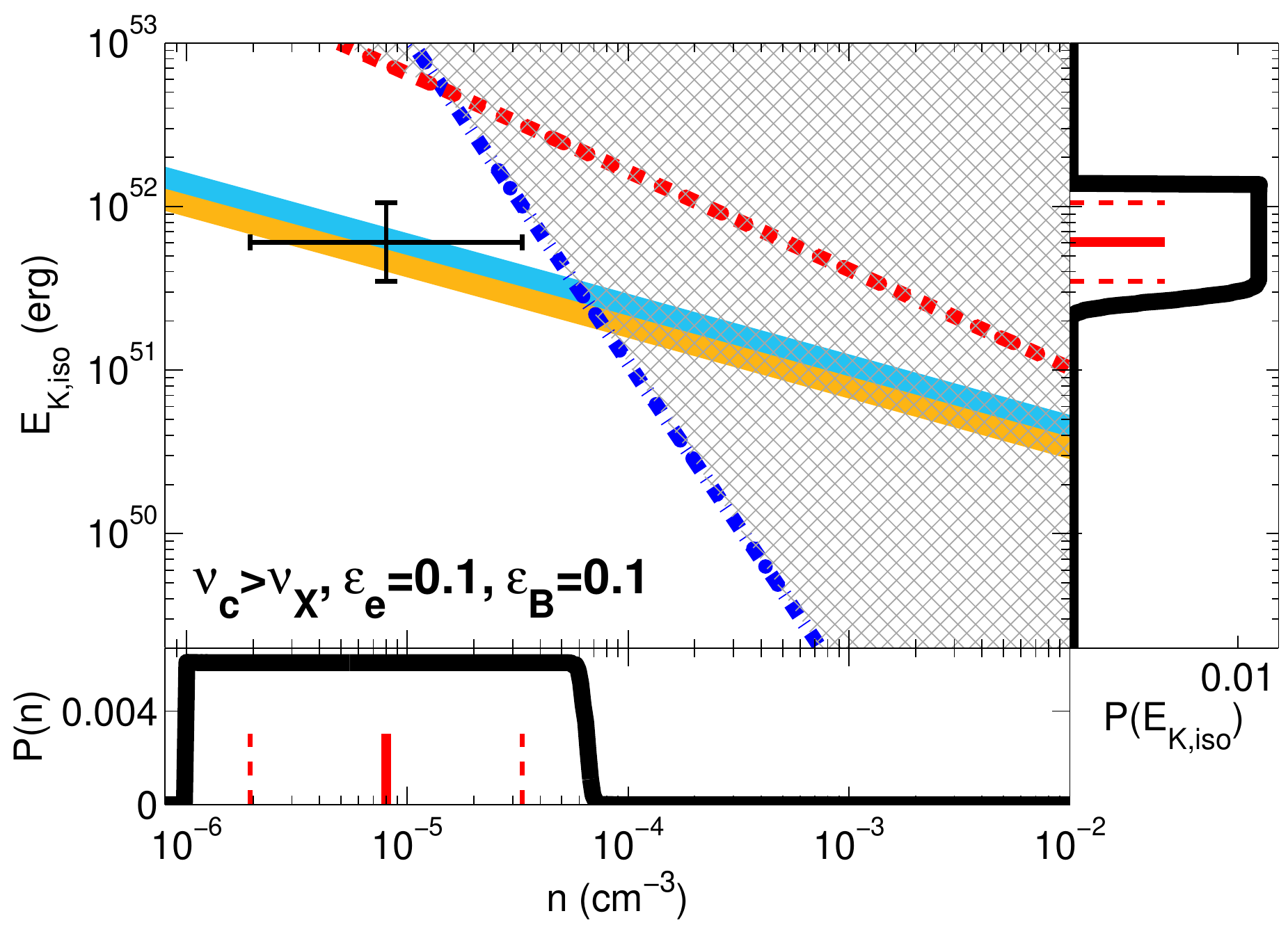}
\includegraphics*[width=0.5\textwidth,clip=]{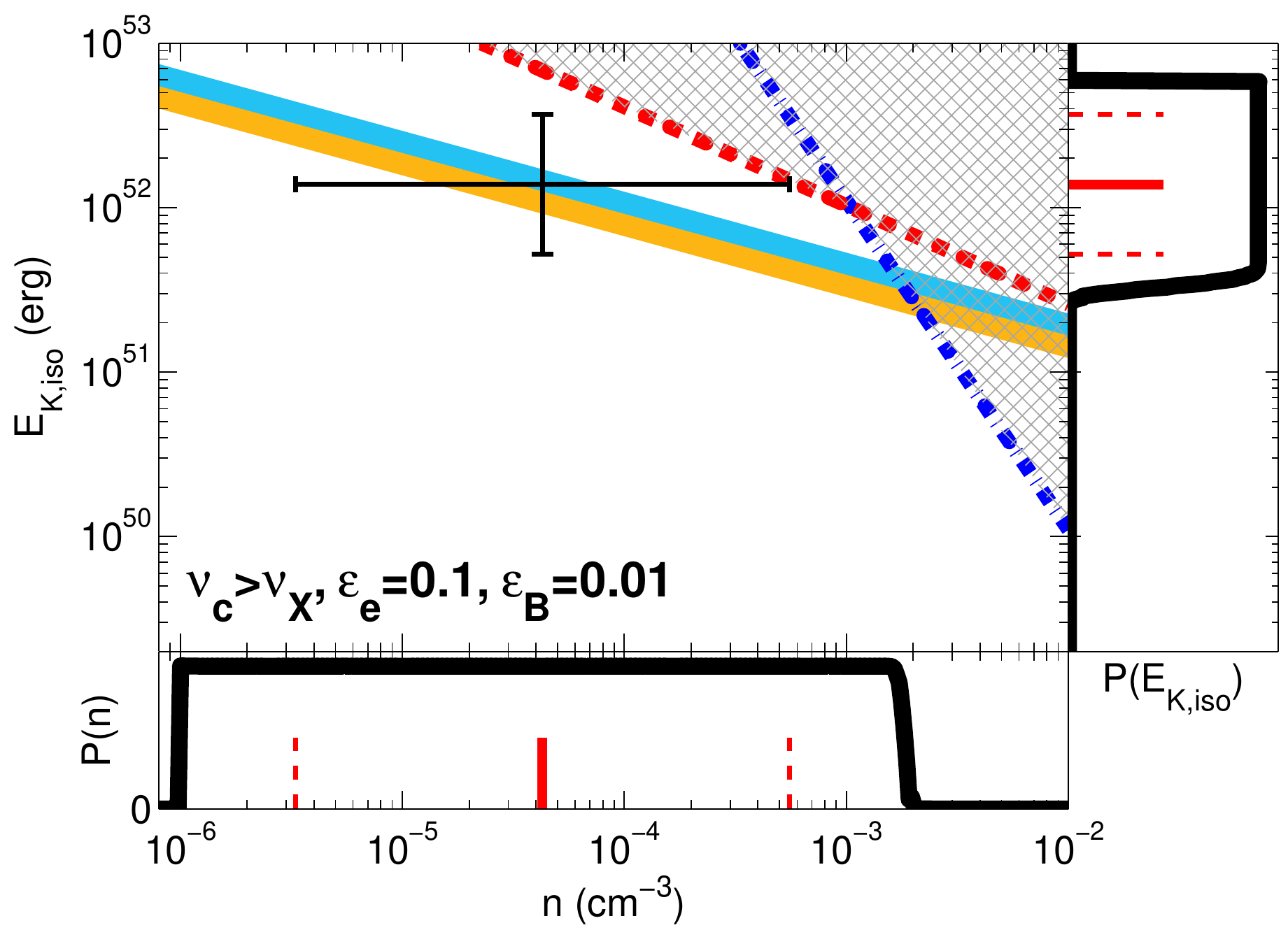} 
\end{minipage}
\vspace{-0.05in}
\caption{Isotropic-equivalent kinetic energy versus circumburst density for \grb, assuming fiducial values for the microphysical parameters of $\epsilon_e=\epsilon_B=0.1$ (left) and $\epsilon_e=0.1$, $\epsilon_B=0.01$ (right). In each panel, the X-rays (light blue), optical (orange) and radio (red) provide independent constraints on the parameter space. Measurements are shown as solid regions, where the width of the region corresponds to the $1\sigma$ uncertainty, while upper limits are denoted as dashed or dot-dashed lines. Setting the cooling frequency to a minimum value of $\nu_{c,{\rm min}}=2.4 \times 10^{18}$~Hz (10~keV) provides an additional constraint (dark blue dot-dashed line). The regions of parameter space ruled out by the observations are denoted (grey hatched regions). The median solution and $1\sigma$ uncertainty is indicated by the black cross in each panel. The joint probability distributions in $n$, with an imposed lower bound of $n_{\rm min}=10^{-6}$~cm$^{-3}$ (bottom panel), and $E_{\rm K,iso}$ (right panel) are shown. Red lines correspond to the median, and dotted lines are the $1\sigma$ uncertainty about the median.
\label{fig:En}}
\end{figure*}

We utilize the broad-band afterglow observations to constrain the explosion properties and circumburst environment of \grb. We adopt the standard synchrotron model for a relativistic blastwave in a constant density medium \citep{spn98,gs02}, as expected for a non-massive star progenitor. This model provides a mapping from the broad-band afterglow flux densities to the burst physical parameters: the isotropic-equivalent kinetic energy ($E_{\rm K,iso}$), circumburst density ($n$), fractions of post-shock energy in radiating electrons ($\epsilon_e$) and magnetic fields ($\epsilon_B$), and the electron power-law distribution index ($p$), with $N(\gamma)\propto \gamma^{-p}$ for $\gamma \gtrsim \gamma_{\rm min}$, where $\gamma_{\rm min}$ is the minimum Lorentz factor of the electron distribution.

To determine the locations of the synchrotron break frequencies with respect to the X-ray, optical, and radio bands, and thus constrain the burst explosion properties, we calculate the electron power-law index $p$, using a combination of temporal and spectral information. The temporal and spectral power-law indices are given by $\alpha$ and $\beta$, respectively, where $F_{\nu} \propto t^{\alpha}\nu^{\beta}$. To determine $\alpha_X$ and $\alpha_{\rm opt}$, we utilize $\chi^2$-minimization to fit a single power law model to each light curve in the form $F_{\nu} \propto t^{\alpha}$. In this manner, we measure $\alpha_X=-1.07 \pm 0.15$ and $\alpha_{\rm opt}=-1.02^{+0.44}_{-0.55}$ for the X-ray and optical light curves, respectively. Computing $\alpha_{\rm opt}$ and $1\sigma$ uncertanties using the detections alone marginally violates the upper limit at $\delta t \approx 10.70$~days; therefore, we use the observation at $\approx10.70$~days to set the upper limit on the decline rate, resulting in asymmetric uncertainties. The afterglow light curves, along with the best-fit power-law models, are shown in Figures~\ref{fig:optlc} and \ref{fig:xraylc}. From the X-ray spectral fit, we find $\beta_{X} \equiv 1-\Gamma_{\rm X}=-0.67 \pm 0.17$ (Section~\ref{sec:xray} and Table~\ref{tab:xrayobs}).

To determine the location of the cooling frequency, $\nu_c$, to the observing bands, we compare the temporal and spectral indices to the closure relations for a relativistic blastwave in a constant density medium \citep{sph99,gs02}. If $\nu_c < \nu_X$ then the independently-derived values for $p$ from the temporal and spectral indices are inconsistent: $p=2.1 \pm 0.2$ from $\alpha_{X}$, $p=1.3 \pm 0.3$ from $\beta_X$, and $p=2.0 \pm 0.73$ from $\alpha_{\rm opt}$. On the other hand, for $\nu_c > \nu_X$, the values are fully consistent: $p=2.4 \pm 0.2$ from $\alpha_{X}$, $p=2.3 \pm 0.3$ from $\beta_X$, and $p=2.4 \pm 0.7$ from $\alpha_{\rm opt}$. Furthermore, if $\nu_c>\nu_X$, then $\beta_X = \beta_{\rm OX}$, where $\beta_{\rm OX}$ is the optical-to-X-ray spectral index. Indeed, extrapolating the \chandra\ data to the time of the optical afterglow using $\alpha_X$, we calculate $\beta_{\rm OX} \approx -0.61$ which is consistent with the value of $\beta_X=-0.67 \pm 0.17$. Therefore for the rest of our calculations, we utilize a value of $p=2.40 \pm 0.17$ as determined by the weighted average and assume $\nu_c>\nu_X$. We note that this is also equal to the average value for $p$ inferred from the short GRB population \citep{fbm+15}.

We next determine a set of constraints on $n$ and $E_{\rm K,iso}$ based on the X-ray and optical flux densities, radio upper limit, and the condition that $\nu_c>\nu_X$. Using the X-ray flux density, $z=0.1343$, luminosity distance $d_L=1.96 \times 10^{27}$~cm, $\nu_x=2.4 \times 10^{17}$~Hz (1~keV), $\nu_{\rm opt}=4.8 \times 10^{14}$~Hz ($r$-band), and $p=2.4$, we obtain the relation \citep{gs02}

\begin{equation}
 n^{0.5} E_{\rm K,iso,52}^{1.35} \epsilon_{e,-1}^{1.4} \epsilon_{B,-1}^{0.85} \approx (1.6 \pm 0.2) \times 10^{-3}
\label{eqn:x}
\end{equation}

\noindent where $n$ is in units of cm$^{-3}$, $E_{\rm K,iso,52}$ is in units of $10^{52}$ erg and $\epsilon_{e,-1}$ and $\epsilon_{B,-1}$ are in units of 0.1. The $1\sigma$ uncertainty is dominated by the uncertainty in the individual flux measurements. The constraint on $E_{\rm K,iso}-n$ parameter space from the X-ray afterglow is shown in Figure~\ref{fig:En}, where the width of the region represents the $1\sigma$ uncertainty. We calculate an independent constraint from the optical flux density, assuming that the optical frequency is between the peak frequency, $\nu_m$, and the cooling frequency, $\nu_c$, such that $\nu_m<\nu_{\rm opt}<\nu_c$. Thus, the optical and X-ray bands occupy the same spectral regime and have the same dependencies on the physical parameters. We calculate the constraint,

\begin{equation}
 n^{0.5} E_{\rm K,iso,52}^{1.35} \epsilon_{e,-1}^{1.4} \epsilon_{B,-1}^{0.85} \approx (8.3 \pm 1.3) \times 10^{-4},
\label{eqn:o}
\end{equation}

\noindent shown in Figure~\ref{fig:En}. Next, we impose a constraint from the radio band under the assumption that the radio band is between the self-absorption frequency and the peak frequency such that $\nu_a<\nu_{\rm radio}<\nu_m$. Using the radio upper limit and $\nu_{\rm radio}=9.8 \times 10^{9}$~Hz, we obtain the constraint

\begin{equation}
n^{1/2} E_{\rm K,iso,52}^{5/6} \epsilon_{e,-1}^{-2/3} \epsilon_{B,-1}^{1/3} \gtrsim 1.5 \times 10^{-2}.
\label{eqn:rad}
\end{equation}

\noindent where the constraint imposed on the  $E_{\rm K,iso}-n$ is shown in Figure~\ref{fig:En}. Using the fact that $\nu_c>\nu_X$, we can use the relative location of the cooling frequency as a final constraint. Setting the cooling break to a minimum value, $\nu_{c,{\rm min}}=2.4 \times 10^{18}$~Hz (10~keV) at the upper edge of the X-ray band, we calculate the relation

\begin{equation}
n^{-1} E_{\rm K,iso,52}^{-1/2} \epsilon_{B,-1}^{-3/2} \gtrsim 3.0 \times 10^{4}.
\label{eqn:vc}
\end{equation}

\noindent We use Equations~\ref{eqn:x}$-$\ref{eqn:vc} and joint probability analysis (described in \citealt{fbm+15}) to solve for the isotropic-equivalent kinetic energy and circumburst density. For $\epsilon_e=\epsilon_B=0.1$, we calculate $n=8.0^{+25.2}_{-6.1} \times 10^{-6}$~cm$^{-3}$ and $E_{\rm K,iso}=6.1^{+4.5}_{-2.6} \times 10^{51}$~erg (Figure~\ref{fig:En}). For $\epsilon_e=0.1$ and $\epsilon_B=0.01$, we find $n=4.3^{+51.1}_{-3.9} \times 10^{-5}$~cm$^{-3}$ and $E_{\rm K,iso}=1.4^{+2.3}_{-0.9} \times 10^{52}$~erg ($1\sigma$ uncertainties). From a comparison of the optical and X-ray afterglow emission, we find no evidence for extinction instrinsic to the burst site or host galaxy (e.g., $A_V^{\rm host}=0$). We note that the values for the kinetic energy and density differ slightly from our previous analysis which included \grb\ \citep{fbm+15} due to a refinement of the optical and X-ray fluxes.

Finally, we calculate the isotropic-equivalent $\gamma$-ray energy, by $E_{\gamma, {\rm iso}}=4\pi d_L^2(1+z)^{-1} f_{\gamma}$~erg. Using the \swift/BAT $15-150$~keV fluence and a bolometric correction factor of $5$ to correspond to a wider $\gamma$-ray energy range of $\approx 10-1000$~keV, we calculate $E_{\gamma, {\rm iso}}=(1.3 \pm 0.3) \times 10^{49}$~erg. 

\subsection{Jet Opening Angle}

We can use the temporal evolution of the afterglow to constrain the jet opening angle. We have no observations of the X-ray afterglow at $\delta t \lesssim 8$~days and therefore no information on the early decline rate. However, the optical and X-ray bands are on the same spectral slope (Section~\ref{sec:sed}) and their afterglows should exhibit the same temporal decline. Indeed, the optical afterglow at $\delta t \approx 1.7-2.7$~days has a decline rate of $\alpha_{\rm opt} \approx -1.02$, matching the X-ray decline rate at $\delta t \approx 7.9-39.7$~days of $\alpha_X \approx -1.07$ within the $1\sigma$ uncertainties. Thus, the combination of the early optical data and the late-time X-ray data implies that the afterglow is on a single power-law decline over $\delta t \approx 1.7-39.7$~days. Since jet collimation is predicted to produce a temporal ``break'' in the light curve (``jet break''; \citealt{sph99}), the single power-law decline can be used to place a lower limit on the opening angle assuming on-axis orientation, as off-axis observing angles could disguise jet breaks \citep{vm12,vm13}. The late-time X-ray observations, in conjunction with the energy, density, and redshift, can be converted to a jet opening angle, $\theta_j$, using \citep{sph99,fks+01}

\begin{equation}
\theta_j=9.5\,t_{j,{\rm d}}^{3/8}(1+z)^{-3/8}E_{\rm K,iso,52}^{-1/8}n_0^{1/8} \text{ deg},
\label{eqn:jb}
\end{equation}

\noindent where $t_{j,{\rm d}}$ is in days. Using $t_{j,{\rm d}}>39.68$~days which corresponds to the last X-ray observation (Table~\ref{tab:xrayobs}), and using the values for the isotropic-equivalent kinetic energy and circumburst density as determined in Section~\ref{sec:sed}, we calculate a lower limit on the opening angle of $\theta_j\gtrsim 9^{\circ}$.

\subsection{Limits on a Supernova or Kilonova}
\label{sec:kn}

\begin{figure}
\centering
\hspace{-0.18in}
\includegraphics*[angle=0,width=0.5\textwidth]{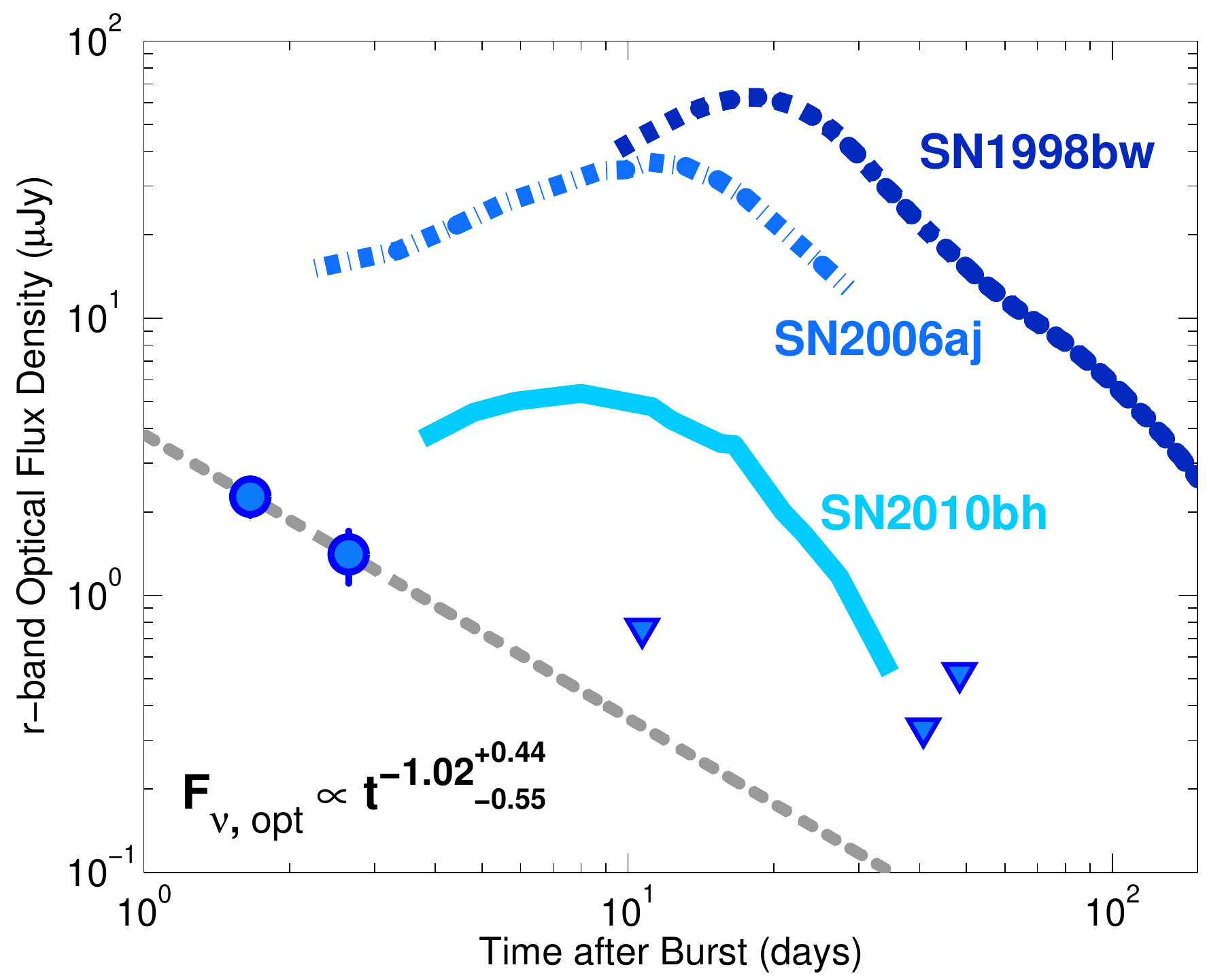}
\caption{Optical $r$-band afterglow light curve of \grb\ (blue circles and triangles). Error bars correspond to $1\sigma$ confidence and triangles denote $3\sigma$ upper limits. The best-fit power-law model (grey dashed line) is characterized by a temporal decay of $\alpha_{\rm opt} \approx -1.02$. Also
shown are the optical light curves of three supernovae which represent the range of luminosities observed for GRB-SN: GRB-SN\,1998bw (dark blue dotted line; \citealt{gvv+98,csc+11}), GRB-SN\,2006aj (blue dot-dashed line; \citealt{mha+06}), and GRB-SN\,2010bh (light blue line; \citealt{ogs+12}); the GRB-SN light curves have all been corrected for extinction and redshifted to $z=0.1343$. The optical limits are a factor of $\approx 6-60$ below the luminosities of known GRB-SNe, ruling out the presence of a SN associated with \grb\ to deep limits.
\label{fig:optlc}}
\end{figure}

We utilize our observations at $\delta t \gtrsim 10$~days to place limits on the presence of supernova or kilonova emission associated with \grb. First, we compare our optical $r$-band observations between $\delta t \approx 10-50$~days to the light curves of known supernovae associated with long-duration GRBs to place limits on supernova emission associated with \grb. We utilize optical light curves of three supernovae which represent the range of luminosities observed for GRB-SN: GRB-SN\,1998bw \citep{gvv+98,csc+11}, GRB-SN\,2006aj \citep{mha+06}, and GRB-SN\,2010bh \citep{ogs+12}, redshifted to $z=0.1343$ (Figure~\ref{fig:optlc}). At $\delta t \approx 10.7$~days, the limit on the optical flux density from \grb\ is $F_{\nu,{\rm opt}}\lesssim 0.76\,\mu$Jy, a factor of $\approx 6-60$ below the expected brightness of GRB-SNe (Figure~\ref{fig:optlc}). Similarly, the limits at $\delta t \approx 40.7$ and $48.4$ days are a factor of $\approx 30-65$ below the expected luminosity of GRB-SN\,1998bw (Figure~\ref{fig:optlc}). Thus, we can rule out the presence of a supernova associated with \grb\ to deep limits.

\begin{figure*}
\begin{minipage}[c]{\textwidth}
\tabcolsep0.0in
\includegraphics*[width=0.247\textwidth,trim={1.2cm 0 4.2cm 0},clip=]{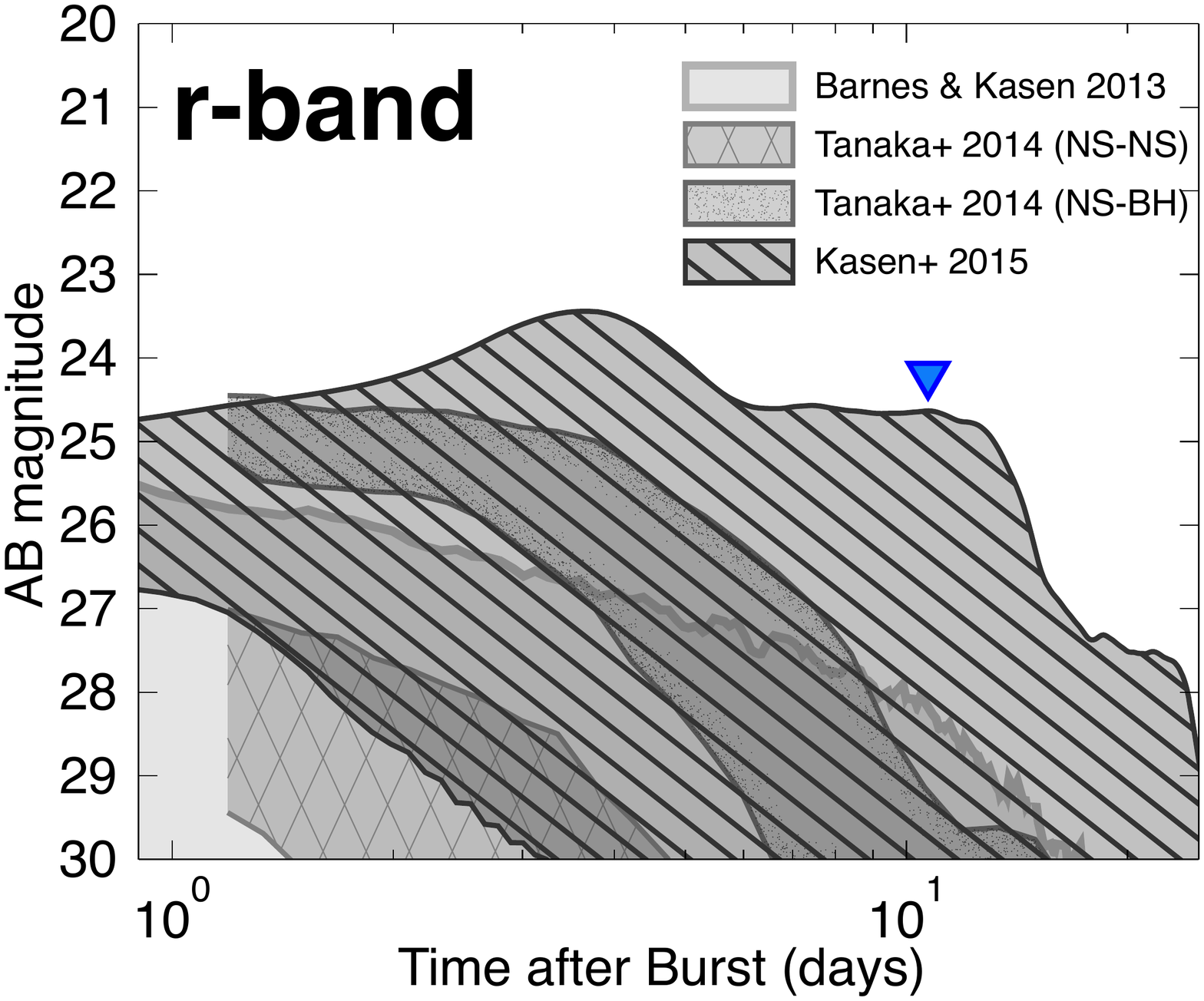}
\includegraphics*[width=0.247\textwidth,trim={1.2cm 0 4.2cm 0},clip=]{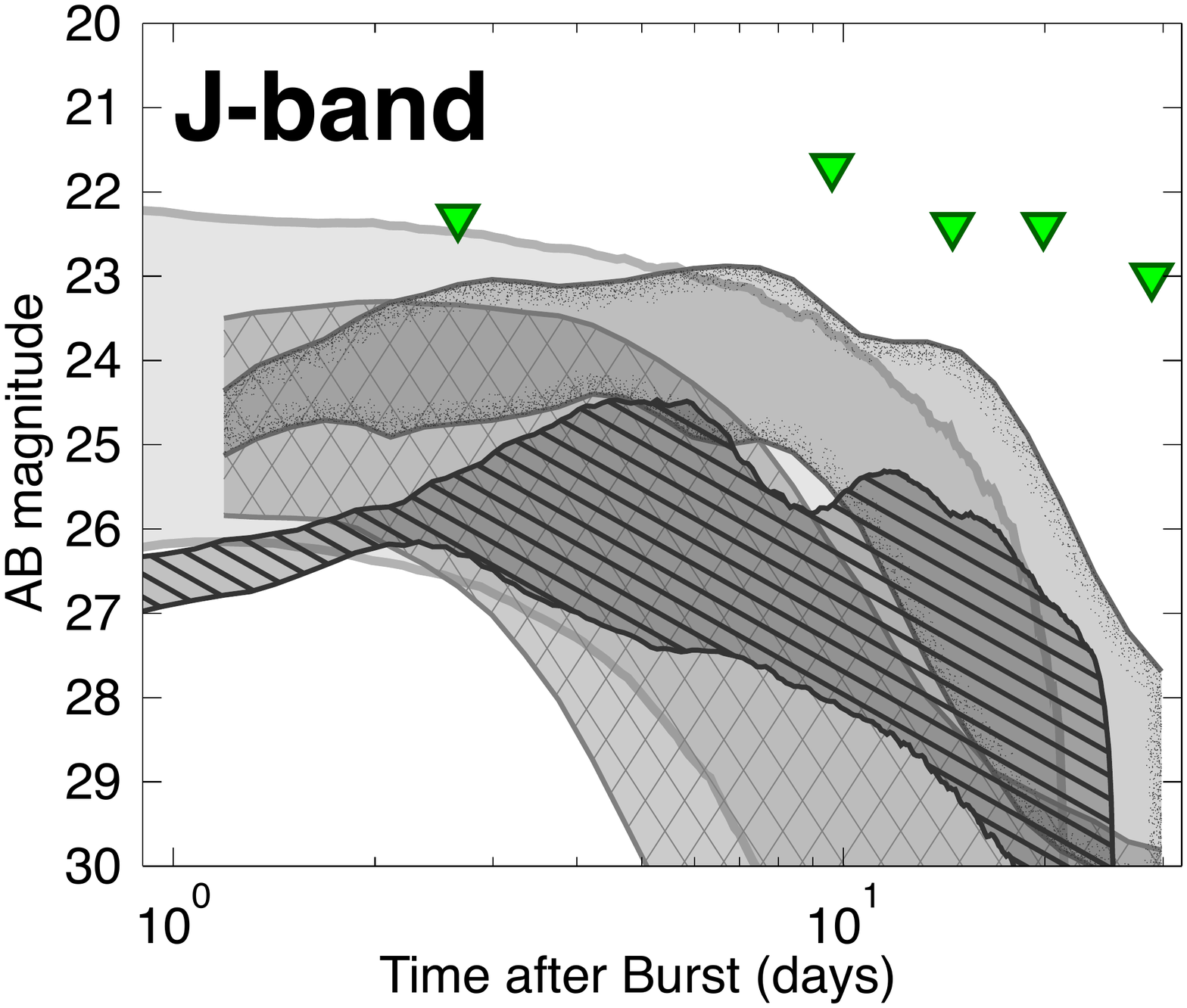} 
\includegraphics*[width=0.247\textwidth,trim={1.2cm 0 4.2cm 0},clip=]{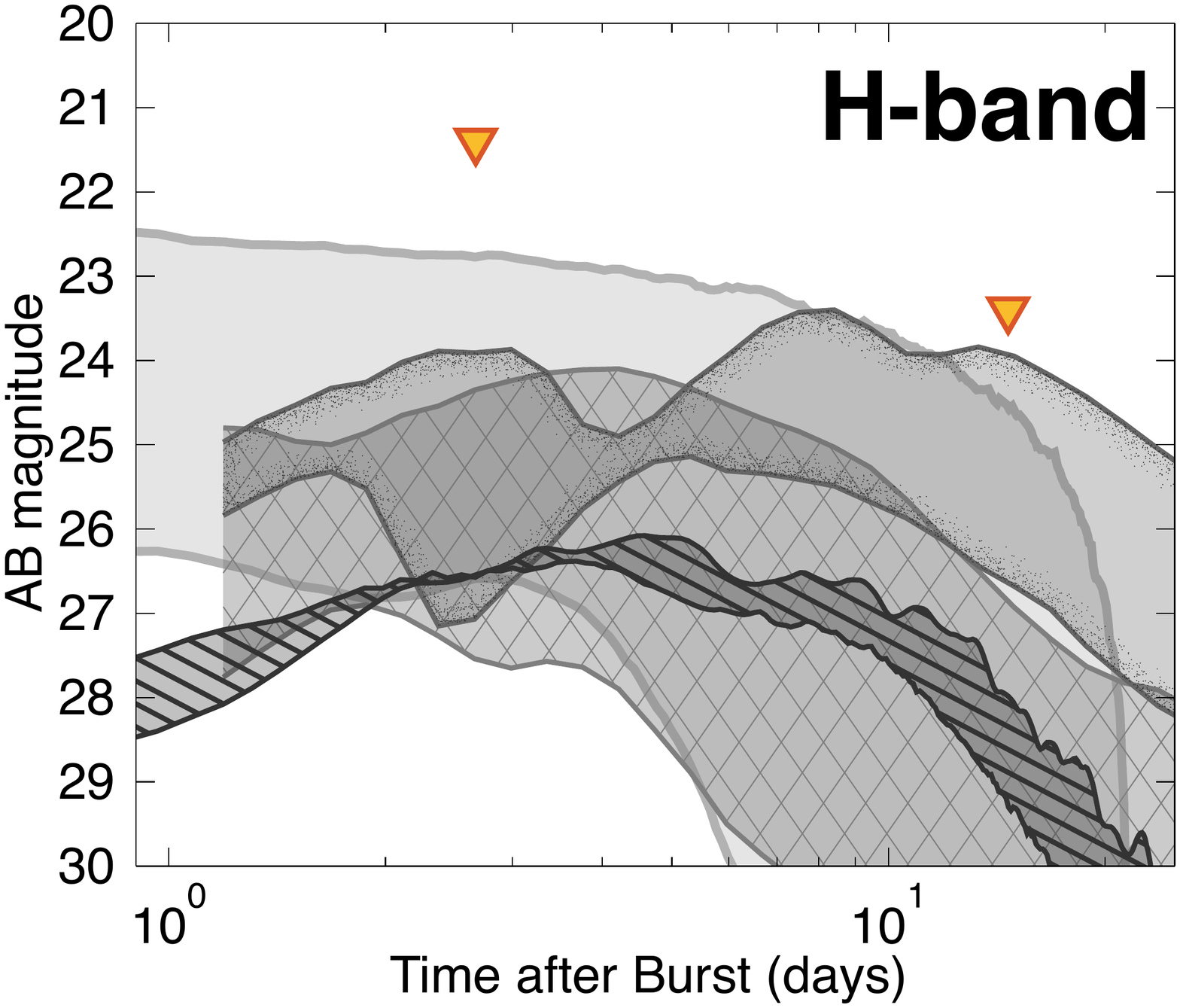} 
\includegraphics*[width=0.247\textwidth,trim={1.2cm 0 4.2cm 0},clip=]{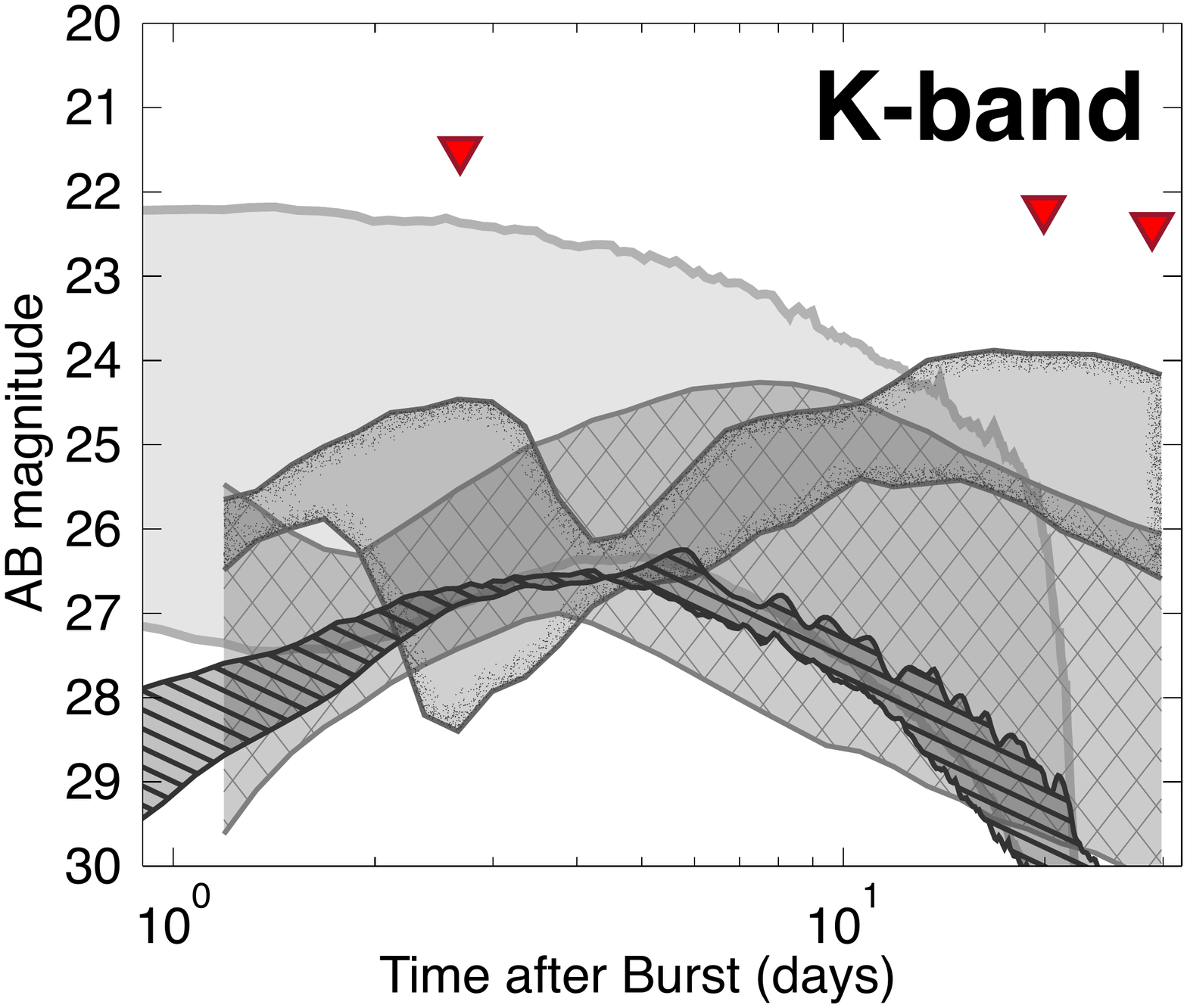}
\end{minipage}
\caption{Optical $r$-band and near-IR $JHK$-band observations of \grb\ between $\delta t \approx 2.6-30$~days, where triangles denote $3\sigma$ upper limits. Also shown are four sets of kilonova models in grey regions: spherically symmetric ejecta from NS-NS mergers (solid; \citet{bk13}), ejecta from NS-NS mergers (crossed) and NS-BH mergers (speckled; \citealt{hkt+13,th13,thk+14}), and outflows from disc winds surrounding a long-lived remnant (single stripes; \citealt{kfm15}). The grey regions represent the full range of light curves considered in each set, assuming pole-on orientation and $r$-process element opacities.
\label{fig:kn}}
\end{figure*}

To place limits on emission from a kilonova associated with \grb, we compare our deep optical and near-IR observations between $\delta t \approx 10.7-28.9$~days to kilonova models from the literature. We note that the available models truncate at $\delta t > 30$~days. The deepest optical limits during this time range are $r_{\rm AB}\gtrsim 24.2$~mag at $\delta t \approx 10.7$~days and $Y_{\rm AB} \gtrsim 23.6$ at $\delta t \approx 14.7$~days. Similarly, the deepest near-IR limits in this time range are $J_{\rm AB} \gtrsim 23.0$~mag, $H_{\rm AB} \gtrsim 23.4$~mag, and $K_{\rm AB} \gtrsim 22.4$~mag (Table~\ref{tab:obsag} and Figure~\ref{fig:kn}).

We consider emission from four sets of kilonova models: (i) spherically symmetric ejecta from NS-NS mergers for a range of ejecta velocities ($0.1-0.3c$) and ejecta masses ($10^{-3}-10^{-1}~M_{\odot}$; \citealt{bk13}), (ii) dynamical ejecta from NS-NS mergers considering ``soft'' and ``stiff'' equations of state for the merger components \citep{hkt+13,th13}, (iii) NS-BH mergers considering ``soft'' and ``stiff'' equations of state for the neutron star and varying spin for the black hole \citep{thk+14}, and (iv) winds from an accretion disc surrounding a long-lived central compact remnant for remnant lifetimes of 0~ms to $\infty$ \citep{kfm15}. We utilize models that have employed $r$-process element (as opposed to iron-like) opacities, where model (i) assumes opacities calculated from a few representative lanthanide elements while models (ii)-(iii) incorporate all of the $r$-process element opacities. For models (ii)-(iv), we select the pole-on orientation light curves. For each observing band, we use models in the corresponding rest-frame band (e.g., $\lambda_{\rm rest}=\lambda_{\rm obs}(1+z)^{-1}$), convert the provided light curves to apparent magnitude, and shift the light curves to the observer frame. The models, along with the late-time optical and near-IR limits are shown in Figure~\ref{fig:kn}. Although these observations place among the deepest limits on kilonova emission following a short GRB to date (see Section~\ref{sec:kiloconstraints}), they do not constrain the brightest models. This demonstrates the difficulty of placing constraints on kilonovae associated with short GRBs at cosmological redshifts based on the current era of kilonova models.  

\section{Host Galaxy Properties}

\begin{figure}
\centering
\hspace{-0.19in}
\includegraphics*[angle=0,width=0.5\textwidth]{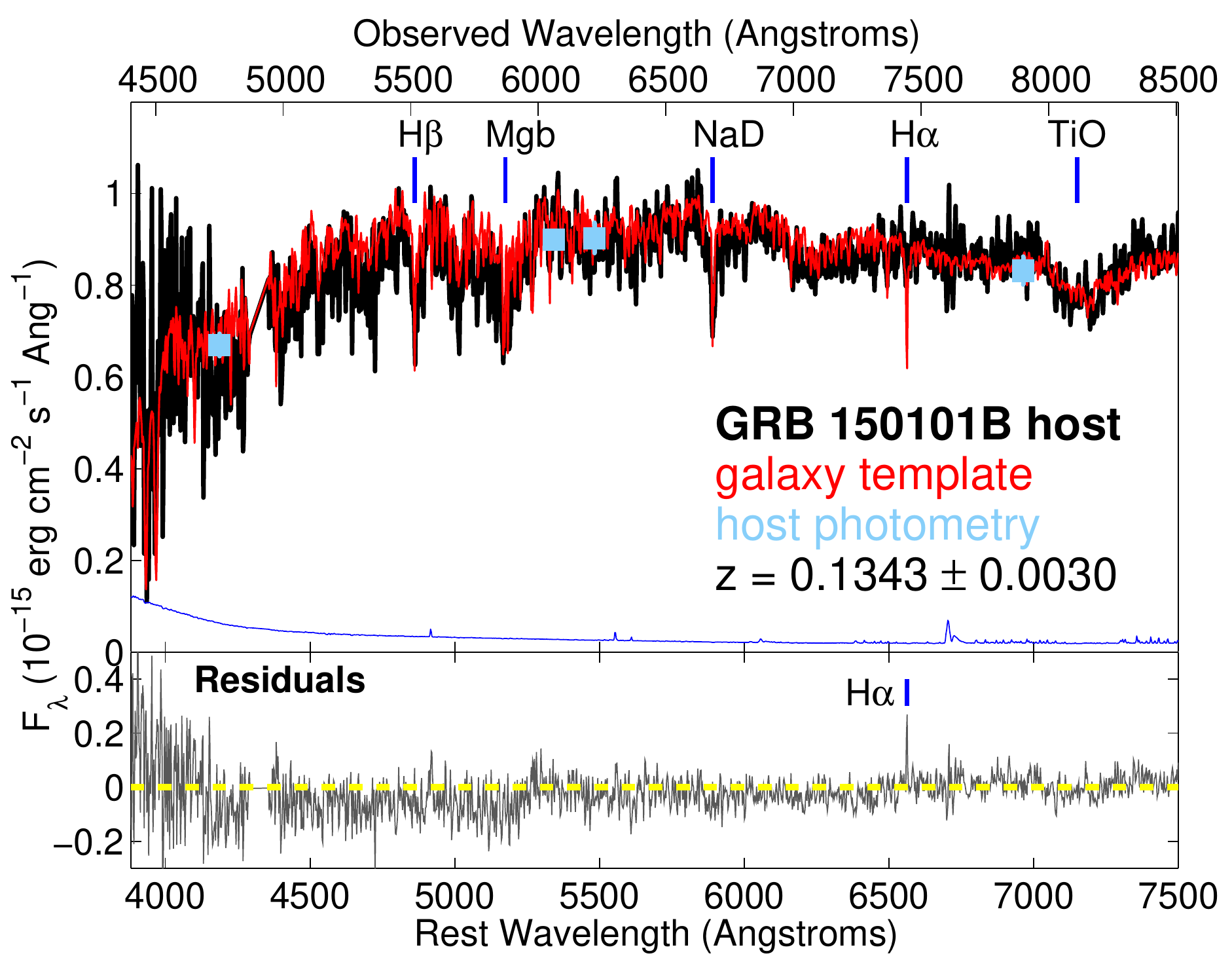}
\caption{{\it Top:} Magellan/Baade IMACS spectrum of the early-type host galaxy of \grb, binned with a 3-pixel boxcar (black: data; blue: error spectrum), and normalized to the flux of the host galaxy in the $gri$-bands and {\it HST}/F606W as determined from aperture photometry (light blue squares). Also shown is an SSP template (red; \citealt{bc03}) with a stellar population age of $2.5$~Gyr at a redshift of $z=0.1343 \pm 0.0030$.  Fits are performed on the unbinned data, and the chip gap ($\lambda_{\rm rest} \approx 4293-4355$\AA) is excluded from the fit. The locations of the Balmer absorption lines, Mgb$\lambda5175$, NaD$\lambda5892$, and TiO$\lambda7050$ are labelled. {\it Bottom:} Subtraction of the galaxy template from the \grb\ host spectrum, revealing emission at the location of H$\alpha\lambda6563$ which indicates the presence of an underlying AGN.
\label{fig:spec}}
\end{figure}

\subsection{Redshift}

To determine the host galaxy's redshift, we fit the IMACS spectrum over the wavelength range of $4400-8500$ \AA\ with simple stellar population (SSP) spectral evolution models at fixed ages ($\tau=0.64-11$~Gyr) provided as part of the GALAXEV library \citep{bc03}; at wavelengths outside this range, the signal-to-noise is too low to contribute significantly to the fit. We use $\chi^{2}$-minimization with redshift as the single free parameter, and perform the fit on the unbinned data. The resulting best-fit model is characterized by a redshift of $z=0.1343 \pm 0030$, determined primarily by the location of the main absorption features of H$\beta$, Mgb$\lambda5175$ and NaD$\lambda5892$ (Figure~\ref{fig:spec}), and a stellar population age of $2.5$~Gyr ($\chi^{2}_{\nu}\approx5.0$ for $1471$ degrees of freedom) assuming solar metallicity. If we allow models to deviate from solar metallicity, adequate fits are also found for stellar population ages of $1.4$ and $5$~Gyr, while poorer fits are found for SSPs with younger or older ages. Due to the deep absorption features, lack of emission lines, and old inferred stellar population age, we classify this host as an early-type galaxy. The shape of the spectrum does not require any intrinsic extinction which is consistent with the results from the afterglow observations. We note that this redshift is fully consistent with the reported redshift from the VLT \citep{lhw+15}.

\subsection{Stellar Population Properties}
\label{sec:ssp}

While the SSP templates provide adequate matches to several of the main absorption features, a strong H$\alpha$ absorption feature is notably absent (Figure~\ref{fig:spec}). Two possible explanations are emission from the underlying AGN (c.f., Section~\ref{sec:agn} and \citealt{xfw+16}) or star formation from a younger stellar population, both of which would cause the H$\alpha$ absorption to be ``filled in''.

To assess these contributions, we subtract the $2.5$~Gyr template from the host spectrum (Figure~\ref{fig:spec}). We find a clear emission feature at the location of H$\alpha$ with an integrated flux of $F_{H\alpha}\approx 1.3 \times 10^{-15}$~erg~s$^{-1}$~cm$^{-2}$, or  $L_{H\alpha}\approx 5.5 \times 10^{40}$~erg~s$^{-1}$ at $z=0.1343$. To help distinguish between an AGN and star formation, we use archival $1.4$~GHz observations as part of the NRAO VLA Sky Survey \citep{ccg+98}, where the host galaxy is detected with $F_{\nu,{\rm 1.4 GHz}}\approx10.2$~mJy. Using standard relations between radio luminosity, $H\alpha$ emission, and star formation \citep{ktc94,yc02,mcs+11}, we find that if the radio emission is due to star formation, the expected H$\alpha$ luminosity is $\approx 500$ times above the observed value. A large SFR is also in contradiction with the early-type galaxy spectrum and lack of emission lines; thus, star formation is not playing a large role. Instead, the H$\alpha$ emission can naturally explained by the AGN, as narrow H$\alpha$ emission is common in the optical spectra of a wide variety of AGN (e.g., \citealt{vv00}). Since H$\alpha$ is contaminated by the AGN, we use the observed H$\alpha$ flux to set an upper limit on the SFR. Using standard relations from \citet{ktc94}, we calculate SFR $\lesssim 0.4$\,M$_{\odot}$~yr$^{-1}$ for the host galaxy of \grb. We note that we do not detect any continuum emission or spectral features at the position of the afterglow. In particular, the lack of emission lines indicates that there is no ongoing star formation at the location of \grb.

To determine the stellar mass and stellar population age, we use the Fitting and Assessment of Synthetic Templates (FAST; \citealt{kvl+09}) code to fit the host galaxy optical and near-IR photometry to a stellar population synthesis template. We fix the redshift to $z=0.1343$ and assume solar metallicity, allowing the extinction, stellar population age, and stellar mass to vary. We include all available host photometry in the fit: $grizJK$-bands from ground-based telescopes and F606W/F160W from {\it HST} (Table~\ref{tab:obsag}). We find a best-fit template characterized by a stellar population age of $\tau=2.0^{+0.8}_{-0.7}$~Gyr and stellar mass of $M_*=7.1^{+1.2}_{-2.3} \times 10^{10}~M_{\odot}$ ($1\sigma$ uncertainties; $\chi^{2}_{\nu}=1.5$). Our results are independent of the choice of template library or initial mass function within the $1\sigma$ uncertainties. We find poorer fits when assuming sub-solar or super-solar metallicities. We note that the age derived from the host galaxy photometry is in good agreement with the results from spectroscopic fitting.

We utilize the host spectrum to calculate the expected $B$-band magnitude as none of the observing bands correspond to rest-frame $B$-band. We calculate the average spectral flux over $100$\,\AA\ centered at $\lambda_{\rm rest}=4450$\,\AA\ and find $F_{\lambda}\approx 7.6 \times 10^{-16}$~erg~cm$^{-2}$~s$^{-1}$\,\AA$^{-1}$; this value does not significantly change for assumed widths of $10-500$\,\AA. The apparent rest-frame $B$-band magnitude is therefore $\approx 17.2$~AB mag, corresponding to an absolute magnitude of $M_B=-21.7$, or $L_B \approx 6.9 \times 10^{10}~L_{\odot}$. From a comparison to the local $B$-band luminosity function which gives $M_B^*=-20.2$ \citep{itz+05} , we infer $L_B \approx 4.3L^*$ for the host of \grb.

\subsection{Morphological Properties}

To determine the shape and size of the host galaxy, we use the IRAF/{\tt ellipse} routine to generate elliptical intensity
isophotes and construct one-dimensional radial surface brightness
profiles. We utilize the late-time {\it HST} observations as these provide the highest angular resolution. For each observation, we
allow the center, ellipticity, and position angle of each isophote to
vary.

Using a $\chi^2$-minimization grid search, we fit each profile with a
S\'{e}rsic model given by

\begin{equation}
\Sigma(r)=\Sigma_e\,{\rm exp}\{-\kappa_n[(r/r_e)^{1/n}-1]\},
\label{eqn:sersic}
\end{equation}

\noindent where $n$ is the concentration parameter ($n=1$ is
equivalent to an exponential disk profile, while $n=4$ is the de
Vaucouleurs profile typical of elliptical galaxies), $\kappa_n\approx
2n-1/3+4/405n+46/25515n^2$ is a constant that depends on $n$
\citep{cb99}, $r_e$ is the effective radius, and $\Sigma_e$ is the
effective surface brightness in flux units. We convert $\Sigma_e$ to
units of mag arcsec$^{-2}$, designated as $\mu_e$. In our grid search,
$n$, $r_e$, and $\mu_e$ are the three free parameters. A single
S\'{e}rsic component provides an adequate fit ($\chi^{2}_{\nu} \approx
1.0$) in both filters. The best-fit models are characterized by $n=5.0 \pm 0.1$ and $r_e=9.5 \pm 0.3$~kpc in the optical F606W filter, and $n=4.1 \pm 0.1$ and $r_e=7.2 \pm 0.2$~kpc in the F160W near-IR filter. Both fits demonstrate an elliptical morphology.

\subsection{Location}

We determine the location of \grb\ with respect to its host galaxy center and light distribution. The angular offset is $\delta R = 3.07 \pm 0.03''$ (Section~\ref{sec:optphot}), which translates to a projected physical offset of $7.35 \pm 0.07$~kpc at $z=0.1343$. Using the effective radius of the host galaxy measured in the F606W and F160W filters, the host-normalized offsets are $0.8r_e$ and $1.0r_e$ in the optical and near-IR bands, respectively.

To determine the brightness of the galaxy at the burst location relative to the host light distribution, we calculate the fraction of total light in pixels fainter than the afterglow position (``fractional flux''; \citealt{fls+06}). Since this relies on high angular resolution, we use the {\it HST} observations where the $1\sigma$ uncertainties on the afterglow positions are $\sigma_{\rm F160W}=0.07''$ and $\sigma_{\rm F606W}=0.04''$ as determined by relative astrometry (Section~\ref{sec:hstobs}). Since the afterglow position spans multiple pixels, we take the average brightness among those pixels to be the representative flux of the afterglow position. For each image, we create an intensity histogram of a region centered on the host galaxy and determine a $1\sigma$ cut-off level for the host by fitting a Gaussian profile to the sky brightness distribution (equivalent to a signal-to-noise ratio cut-off of 1). We then plot the pixel flux distribution above the appropriate cut-off level for a region surrounding the host, and determine the fraction of light in pixels fainter than the afterglow pixel. In this manner, we calculate fractional flux values of 0.21 (F606W) and 0.34 (F160W). 

\section{Discussion}

\subsection{Afterglow and Explosion Properties}

We compare the inferred afterglow and burst explosion properties of \grb\ to previous events. With an observed duration of $T_{90}\approx0.012$~sec, \grb\ has one of the shortest durations for an event detected by {\it Swift} or {\it Fermi} to date \citep{lsb+16}. This holds true when comparing its intrinsic, rest-frame duration of $\approx 0.011$~sec to the population of {\it Swift}-detected short GRBs with determined redshifts \citep{lsb+16}. The temporal evolution of the X-ray and optical afterglows are characterized by power-law decline rates that are consistent with the median values for the short GRB population of $\langle \alpha_X \rangle \approx -1.07$ and $\langle \alpha_{\rm opt} \rangle \approx -1.07$ \citep{fbm+15}. The detection of the X-ray afterglow of \grb\ extends to $\delta t\approx 40$~days, corresponding to a rest-frame time of $\delta t_{\rm rest} \approx 35$~days.

\begin{figure}
\centering
\includegraphics*[angle=0,width=0.5\textwidth]{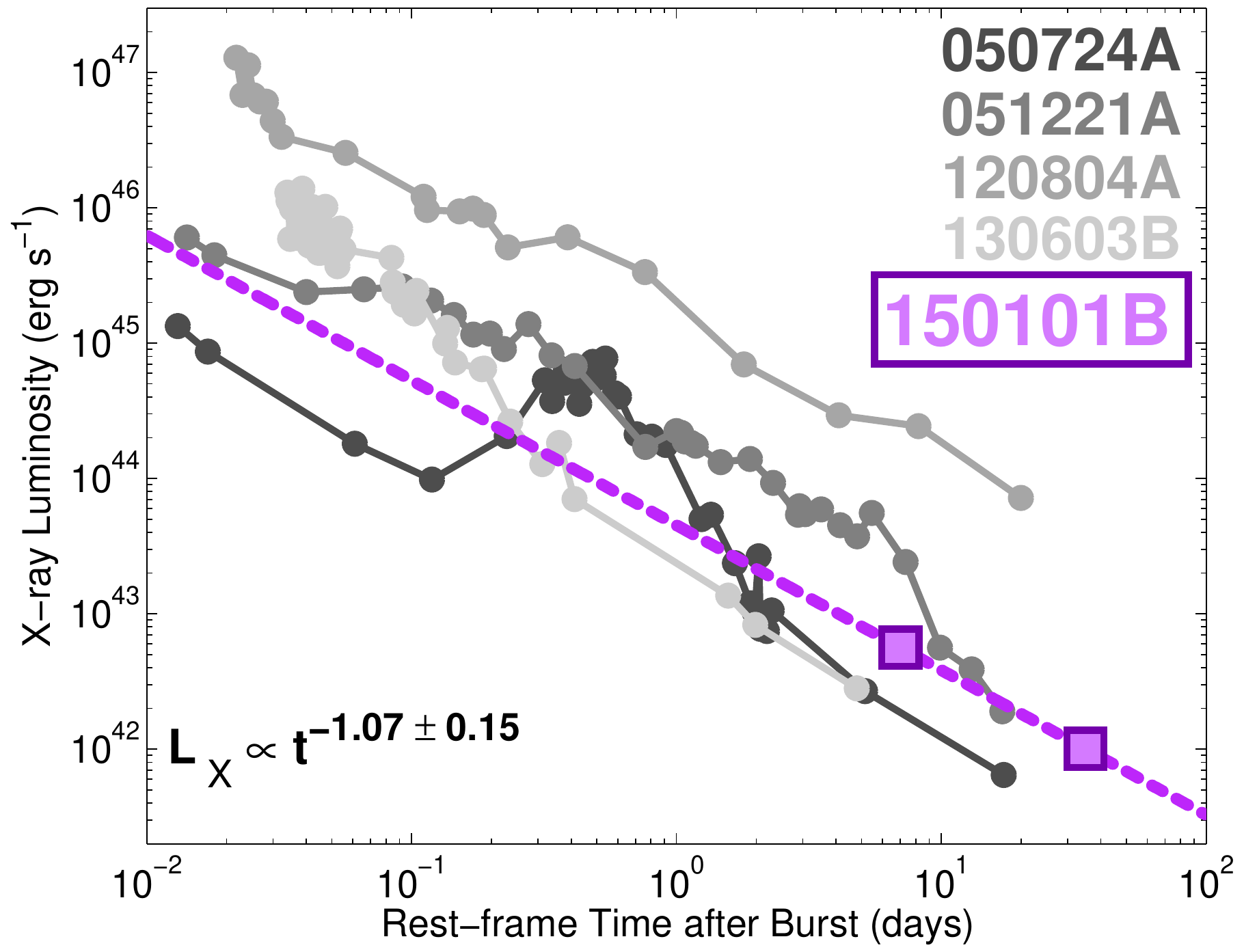}
\caption{X-ray afterglow light curve of \grb\ in the rest-frame of the burst (purple squares). The best-fit power-law model (purple dashed line) is characterized by a temporal decay of $\alpha_{\rm X} = -1.07 \pm 0.15$. Also shown are the X-ray light curves of four short GRBs with late-time X-ray observations and determined redshifts (grey circles): GRB\,050724A \citep{gbp+06}, GRB\,051221A \citep{bgc+06,sbk+06}, GRB\,120804A \citep{bzl+13}, and GRB\,130603B \citep{fbm+14}. The light curves are truncated at $\delta t \lesssim 0.01$~d for clarity, and grey lines connect observations for the same burst. The low redshift of \grb, coupled with the moderate decline rate of $L_X \propto t^{-1.07}$ at late times allows for the detection of the X-ray afterglow to an unprecedented time of $\delta t_{\rm rest} \approx 35$~days.
\label{fig:xraylc}}
\end{figure}

To put \grb\ in the context of other events, we collect X-ray light curves for all \swift\ short GRBs which have multiple detections beyond $\delta t_{\rm rest} \gtrsim 1$~day and determined redshifts. These selection criteria result in four events: GRB\,050724A \citep{gbp+06}, GRB\,051221A \citep{bgc+06,sbk+06}, GRB\,120804A \citep{bzl+13}, and GRB\,130603B \citep{fbm+14}. For \grb\ and each of the previous events, we convert the X-ray flux (0.3-10~keV) to luminosity and shift the light curve to the rest-frame of the burst. The resulting X-ray light curves are shown in Figure~\ref{fig:xraylc}. \grb\ has an X-ray luminosity of $L_X \approx 5.6 \times 10^{42}$~erg~s$^{-1}$ at $\delta t_{\rm rest}\approx 7.0$~days, similar to the X-ray luminosities of previous short GRBs at the same epoch (Figure~\ref{fig:xraylc}). Previously, short GRBs have only been detected to $\delta t_{\rm rest} \approx 20$~days. The low redshift of \grb, coupled with the moderate decline rate of $L_X \propto t^{-1.07}$ at late times allows for the detection of the X-ray afterglow to an unprecedented time of $\delta t_{\rm rest} \approx 35$~days.

We compare our inferred values of the explosion properties for \grb\ to those for the population of \swift\ short GRBs. We calculate median values for the isotropic-equivalent energies of $E_{\gamma,{\rm iso}} \approx 1.3 \times 10^{49}$~erg and $E_{\rm K,iso} \approx (6-14) \times 10^{51}$~erg, where the range is due to uncertainty in the microphysical parameters. From the combination of the X-ray and optical afterglows, we infer a lower limit on the jet opening angle of $\theta_j \gtrsim 9^{\circ}$ for \grb. Using $9^{\circ}$ as the minimum opening angle, we find minimum beaming-corrected energies of $E_{\gamma}\gtrsim 1.6 \times 10^{47}$~erg and $E_{\rm K} \gtrsim (7.4-17) \times 10^{49}$~erg. Compared to \swift\ short GRBs, the isotropic-equivalent $\gamma$-ray energy of \grb\ is one of the lowest values for short GRBs, while the inferred isotropic-equivalent kinetic energy is at the $60-70\%$ level of the population \citep{fbm+15}. Assuming a constant density medium, we also infer an extremely low median circumburst density of $\approx (0.8-4.3) \times 10^{-5}$~cm$^{-3}$ (for $\epsilon_B=0.01-0.1$), which is in the bottom $\approx 1/4$ of all events, and at the median level when compared to short GRBs which originated in elliptical host galaxies \citep{fbm+15}. These low densities are commensurate with predictions for NS-NS mergers, as population synthesis for mergers in large, $10^{11}$~$M_{\odot}$ galaxies predict a significant fraction of systems should occur in environments with gas densities of $\lesssim 10^{-3}$~cm$^{-3}$ \citep{bpb+06}.

\subsection{Host Galaxy Environment: Implications for the Progenitor}

At $z=0.1343$, \grb\ is the closest short GRB with an early-type host galaxy to date. The host association is extremely robust, with a probability of chance coincidence of $P_{cc}\approx 5 \times 10^{-4}$. A single more nearby event\footnote{The short GRB\,061201 has a possible association with a star-forming host galaxy at $z=0.111$, but with a relatively high $P_{cc} \approx 0.1-0.2$ \citep{sdp+07,fbf10}.}, GRB\,080905A, likely occurred in a star-forming host galaxy at $z=0.1218$, with $P_{cc}\approx 0.01$ \citep{rwl+10}. Overall, short GRBs in early-type galaxies comprise $\approx 30\%$ of the population and are detected to $z\sim 1$ \citep{fbc+13}, similar to the host galaxy demographics of Type Ia SNe (e.g., \citealt{slp+06}). Both populations are indicative of older stellar progenitors. In contrast, long GRBs exclusively occur in star-forming galaxies \citep{wbp07,sgl09}, consistent with their massive star progenitors. Furthermore, the lack of associated supernova demonstrates that \grb\ did not form from a massive star progenitor. 

From fitting of the optical to near-IR SED, we infer stellar population properties for the host: a stellar mass of $\approx 7 \times 10^{10}~M_{\odot}$ (log$(M_*/M_{\odot})\approx 10.9$), stellar population age of $\approx 2-2.5$~Gyr (at solar metallicity), star formation rate of $\lesssim 0.4~M_{\odot}$~yr$^{-1}$, and $B$-band luminosity of $\approx 6.9 \times 10^{10}~L_{\odot}$ ($\approx 4.3L^{*}$). These properties are consistent within $2\sigma$ to the values inferred from an independent analysis of the radio to X-ray SED \citep{xfw+16}. The stellar mass and stellar population age are well-matched to the eight previous short GRBs with early-type host galaxies, which have stellar masses of log($M_*/M_{\odot})\approx 10.3-11.8$ and stellar population ages of $\approx 0.8-4.4$~Gyr \citep{lb10}. The host of \grb\ is one of the most luminous early-type host galaxies, surpassed only by GRB\,050509B with $5.0L^*$ \citep{bpp+06}. This is commensurate with its large relative size of $\approx 8$~kpc. Overall, the global host galaxy properties indicate an evolved stellar population and are consistent with those expected for an older stellar progenitor. Notably, this is the first evidence for an AGN in a GRB host galaxy; the paucity of GRB-AGN systems is perhaps not surprising given the rates of low-luminosity AGN and well-studied GRB hosts \citep{xfw+16}.

\begin{figure*}
\begin{minipage}[c]{\textwidth}
\tabcolsep0.0in
\includegraphics*[width=0.5\textwidth,trim={1cm 0 3.5cm 0},clip=]{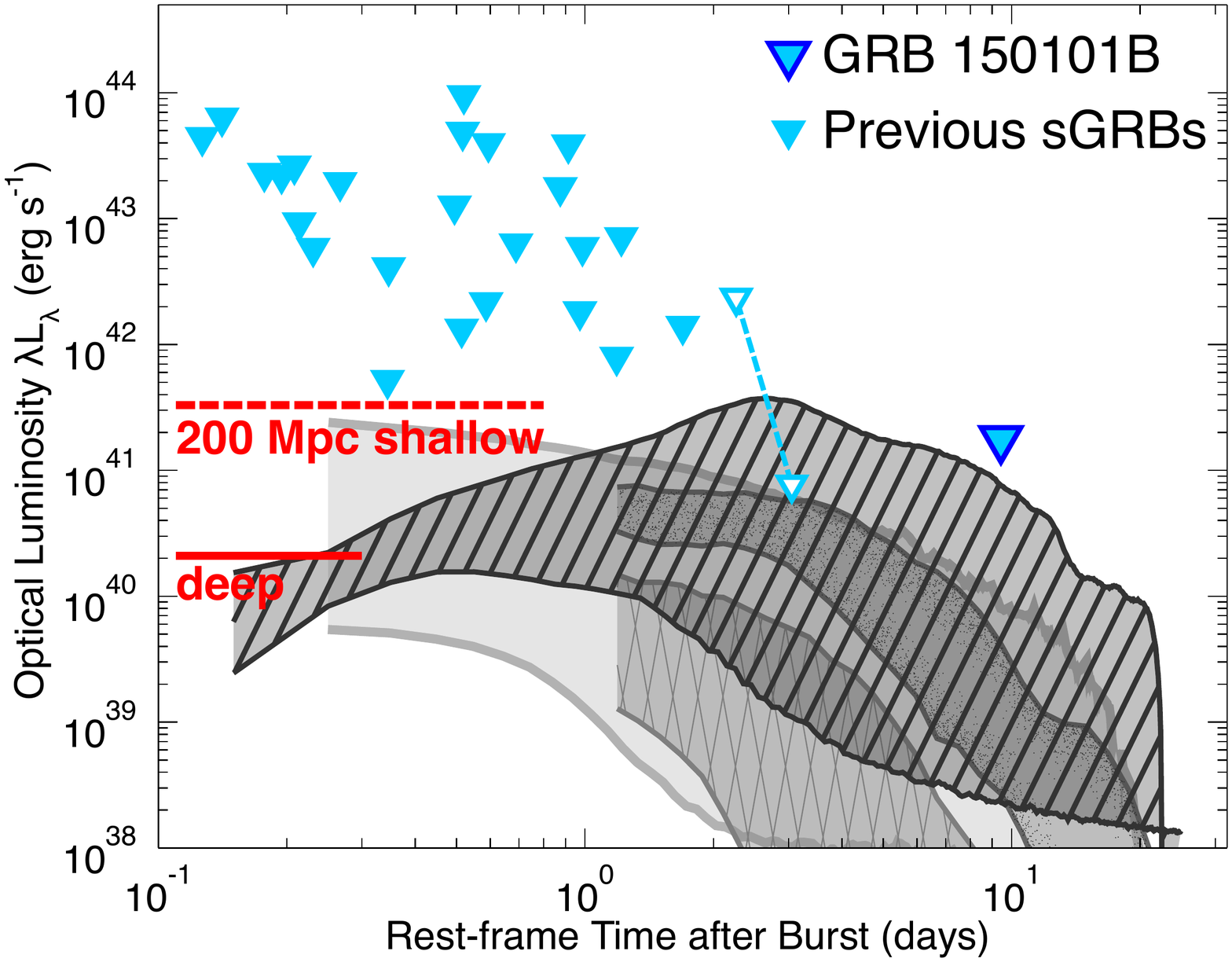}
\includegraphics*[width=0.5\textwidth,trim={1cm 0 3.5cm 0},clip=]{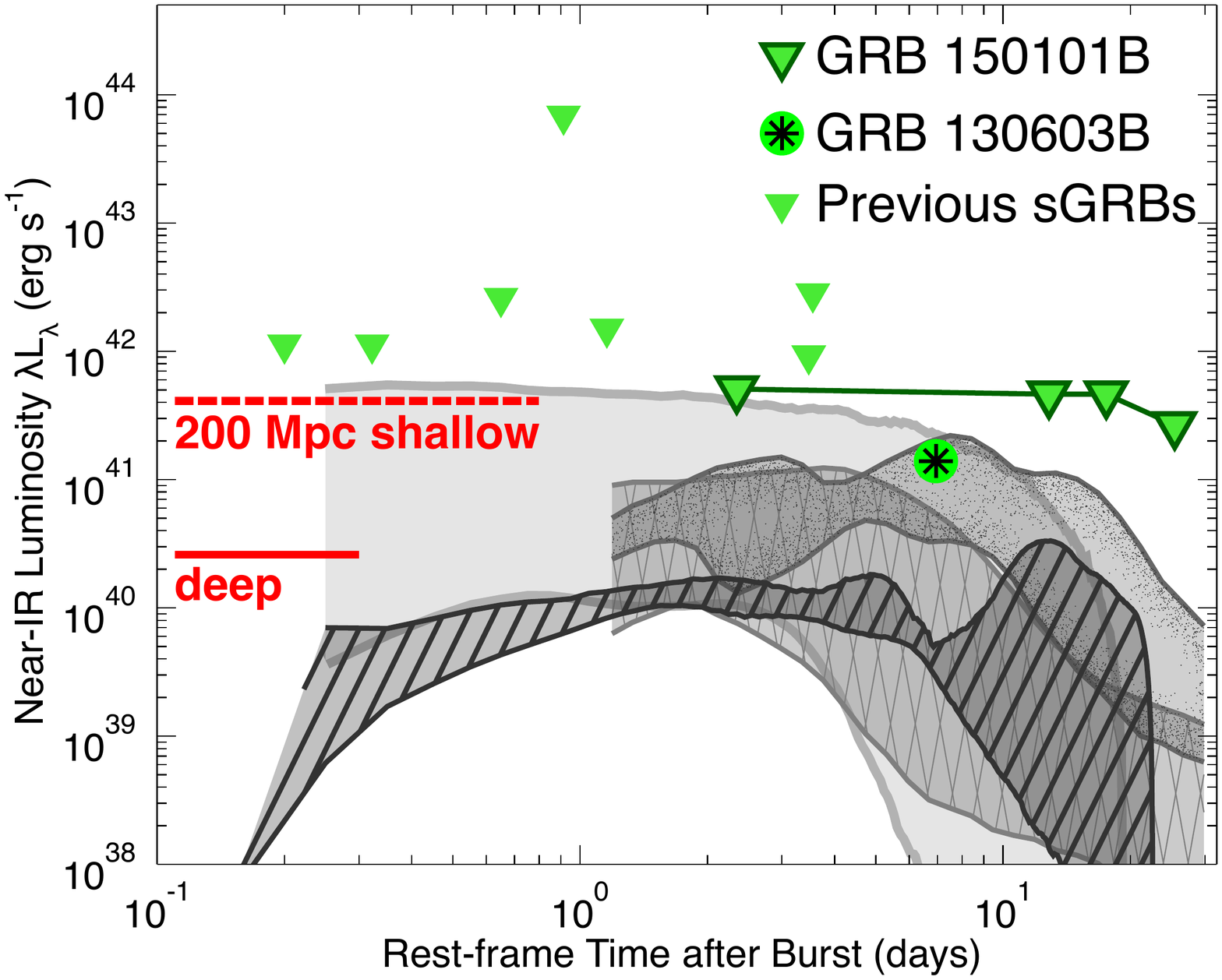} 
\end{minipage}
\vspace{-0.2in}
\caption{Constraints on kilonova emission for \grb\ and all previous short GRBs with observations at $\delta t_{\rm rest} \gtrsim 0.1$~days in the rest-frame optical $r$-band (left) and near-IR $J$-band (right) emission. Triangles denote $3\sigma$ limits, from this work and the short GRB afterglow catalog \citep{fbm+15}. Open triangles denote limits for GRB\,061201 at a tentative redshift of $z=0.111$, and at the median short GRB redshift of $z=0.5$. Also shown is the detection of the near-IR kilonova following GRB\,130603B (black asterisk; \citealt{bfc13,tlf+13}). Grey regions denote four sets of kilonova models in $r$- and $J$-bands (same as in Figure~\ref{fig:kn}). Dashed and solid lines represent shallow ($r_{\rm AB}=21$~mag and $J_{\rm AB}=20$~mag) and deep ($r_{\rm AB}=24$~mag and $J_{\rm AB}=23$~mag) searches following an event at $200$~Mpc. The optical and near-IR limits for \grb\ provide the deepest constraints to date on kilonova emission. A comparison of the existing data to kilonova models demonstrates the difficulty of placing meaningful constraints on kilonova emission with the cosmological sample of short GRBs, based on the current era of models. At a distance of 200~Mpc, shallow searches following gravitational wave events may only be marginally effective in detecting kilonovae. However, deep searches to depths of $\approx 2 \times 10^{40}$~erg~s$^{-1}$ will probe a meaningful range of kilonova models.
\label{fig:kncomp}}
\end{figure*}

Another important diagnostic to evaluate the progenitor is the location of \grb\ with respect to its host galaxy. Indeed, a hallmark signature of NS-NS/NS-BH mergers are natal kicks imparted to the compact objects during their formation (e.g., \citealt{fbb98,fry04}). Coupled with the $\sim$Gyr delay times, this results in a population of events with a range of separations from their hosts \citep{pb02,bpb+06,brf14}. We calculate a kick velocity for \grb\ by assuming the progenitor system formed with the most recent stellar population of $\approx 2-2.5$~Gyr. Using the projected physical offset of $\approx 7.4$~kpc, this implies a minimum kick velocity of $v_{\rm min}\approx 2.9$~km s$^{-1}$ for the progenitor of \grb. However, a more reasonable value for the kick velocity takes into account the host velocity dispersion ($v_{\rm disp}$), such that $v_{\rm kick} \approx \sqrt{v_{\rm min} v_{\rm disp}}$. Assuming $v_{\rm disp} \approx 250$~km s$^{-1}$ as determined from $\approx 10^{11}\,M_{\odot}$ elliptical galaxies \citep{fp99}, we find $v_{\rm kick} \approx 27-30$~km~s$^{-1}$ for \grb. This is on the low end of the inferred natal kick velocities for the eight known Galactic NS-NS binaries which range from $\approx 5-500$~km~s$^{-1}$ \citep{fk97,wkh04,wwk10}. We note that \citet{xfw+16} inferred an older stellar population age of $\approx 5.7 \pm 1.0$~Gyr, which would translate to a lower kick velocity of $v_{\rm kick}\approx 16-20$~km~s$^{-1}$. The fractional flux value of $\approx 20-35\%$ demonstrates that \grb\ occurred on a faint region of its host rest-frame optical light, and is thus weakly correlated with local stellar mass. Furthermore, there is no evidence for ongoing star formation at the position of \grb. These findings are also commensurate with NS/BH kicks. 

\subsection{Constraints on Kilonova Emission}
\label{sec:kiloconstraints}

A predicted signal of NS-NS/NS-BH mergers is transient emission from the radioactive decay of heavy elements produced in the merger ejecta, ($r$-process ``kilonova''; \citealt{lp98,kul05,mmd+10}). The signal is predicted to be dominant in the near-IR bands due to the heavy element opacities \citep{bk13,rka+13,th13}, although models incorporating a long-lived NS remnant predict bluer colors \citep{mf14,kfm15}. A near-IR excess detected with {\it Hubble Space Telescope} ({\it HST}) observations following the short GRB\,130603B was interpreted as kilonova emission and the first direct evidence that short GRBs originate from NS-NS/NS-BH mergers \citep{bfc13,tlf+13}.

For \grb, we place limits of $\approx (2-4) \times 10^{41}$~erg~s$^{-1}$ on kilonova emission (Figure~\ref{fig:kncomp}). To compare these limits to searches for late-time emission following previous short GRBs, we collect all available data from the short GRB afterglow catalog  \citep{fbm+15},  constraining the sample to bursts with upper limits at $\delta t_{\rm rest} \gtrsim 0.1$~days to match the timescale of kilonova light curves. We only include events which have either rest-frame $r$- or $J$-band observations. For 15 bursts with no determined redshift, we assume the median of the short GRB population, $z=0.5$, to convert to luminosity. In addition to \grb, $25$~short GRBs have rest-frame optical limits and seven events have rest-frame near-IR limits. These limits, along with the GRB\,130603B-kilonova detection \citep{bfc13,tlf+13} and four sets of kilonova models (described in Section~\ref{sec:kn}), are displayed in Figure~\ref{fig:kncomp}. We note that since GRB\,061201 has a relatively uncertain association with a galaxy at $z=0.111$, we also display the limit if this burst originated at the median redshift of $z=0.5$ (Figure~\ref{fig:kncomp}).

In the rest-frame optical band, \grb\ has one of the deepest limits on optical kilonova emission to date with $\approx 2 \times 10^{41}$~erg~s$^{-1}$, and the most stringent for a short GRB with a secure redshift. For GRB\,061201, if the true redshift is $z=0.111$, this event has the deepest limit of $\approx 6 \times 10^{40}$~erg~s$^{-1}$. This limit can rule out the optically brightest models which invoke an {\it indefinitely} stable NS remnant \citep{kfm15}, while an assumption of a higher-redshift origin at $z=0.5$ is not stringent enough to place any meaningful constraints (Figure~\ref{fig:kncomp}).

The sample of short GRBs with rest-frame near-IR follow-up is significantly smaller, spanning a range of $\approx 10^{42}-10^{44}$~erg~s$^{-1}$ with most limits clustered at $\approx (0.8-3) \times 10^{42}$~erg~s$^{-1}$. Thus, with constraints of $\approx (2-4) \times 10^{41}$~erg~s$^{-1}$, \grb\ has the deepest limit on the luminosity of a near-IR kilonova (Figure~\ref{fig:kncomp}). For comparison, the detection of the near-IR kilonova following GRB\,130603B had a luminosity of $\approx 1.5 \times 10^{41}$~erg~s$^{-1}$, which mapped to an ejecta mass and velocity of $\approx 0.03-0.08M_{\odot}$ and $\approx 0.1-0.3c$ \citep{bfc13,tlf+13}. In the case of \grb, optical observations of comparable depth at earlier epochs of $\delta t_{\rm rest} \approx 2-5$~days or deeper near-IR observations at $\delta t_{\rm rest} \lesssim 10$~days would have helped to confirm or rule out the brightest kilonova models (Figure~\ref{fig:kncomp}). This demonstrates the difficulty of performing effective kilonova searches following short GRBs based on the current era of kilonova models, and the necessity of more sensitive instruments (e.g., space-based facilities or $\sim$30-m ground-based telescopes) in this effort.  

A more promising route to kilonova detection is searches following NS-NS/NS-BH mergers detected by the Advanced LIGO-VIRGO (ALV) network, which is expected to detect such systems to 200~Mpc at design sensitivity \citep{aaa+13}. Due to their isotropic emission and distinct red color, kilonovae are expected to be premier counterparts in the optical and near-IR bands \citep{mb12}. Given current models, shallow searches that reach depths of $r_{\rm AB}=21$~mag and $J_{\rm AB}=20$~mag, corresponding to $\approx$few~$\times 10^{41}$~erg~s$^{-1}$ at 200~Mpc, will only be marginally effective in detecting additional kilonovae (Figure~\ref{fig:kncomp}). However, deep searches reaching depths of $r_{\rm AB}=24$~mag and $J_{\rm AB}=23$~mag, corresponding to $\approx 2 \times 10^{40}$~erg~s$^{-1}$ will be sensitive to a more meaningful range of existing kilonova models, and thus improve the chance of detection (Figure~\ref{fig:kncomp}).

\section{Conclusions}

We presented observations of the broad-band afterglow and host galaxy environment of the short \grb. The detection of the X-ray and optical afterglows provided sub-arcsecond localization, pinpointing \grb\ to an early-type host galaxy at $z=0.1343$. At this redshift, \grb\ is the closest short GRB with an early-type host galaxy to date. In addition, the stellar population properties are well-matched to those of early-type hosts of previous short GRBs.

The lack of associated supernova demonstrates that \grb\ did not originate from a massive star progenitor. Furthermore, the old $\approx 2-2.5$~Gyr stellar population, combined with the location of \grb\ in a faint region of its host rest-frame optical light are fully consistent with predictions for a NS-NS/NS-BH progenitor. Using the projected physical offset of $\approx 7.4$~kpc and stellar population age as a proxy for delay time, the inferred natal kick velocity is $\approx 27-30$~km~s$^{-1}$, consistent with inferred kick velocities of Galactic double pulsars.

Optical and X-ray observations over $\approx 2-40$~days allow for the determination of the burst basic explosion properties. The isotropic-equivalent $\gamma$-ray energy is one of the lowest inferred for a short GRB, while the kinetic energy is comparable to those of previous short GRBs. The circumburst density is also comparatively low, in the bottom $\approx 25\%$ of all events, while at the median level when compared to events that originated in early-type galaxies. The inferred opening angle of $\gtrsim 9^{\circ}$ is further support that short GRBs have wider jets than those of their long-duration counterparts.

Late-time optical and near-IR observations place limits on the luminosity of optical and near-IR kilonova emission following \grb\ of $\approx (2-4) \times 10^{41}$~erg~s$^{-1}$, among the most constraining limits ever placed on kilonova emission following a short GRB. A comparison of existing observations following short GRBs to current kilonova models demonstrates the difficulty of performing meaningful kilonova searches for the cosmological sample of short GRBs. This comparison also highlights the importance of observations following short GRBs at $\lesssim 10$~days with space-based facilities or upcoming extremely large-aperture telescopes in detecting kilonovae at cosmological distances.

At design sensitivity, the ALV network is expected to provide detections of NS-NS/NS-BH mergers to $200$~Mpc \citep{aaa+13}. Given current kilonova models, deep optical and near-IR searches to depths of $\approx23-24$~AB~mag are necessary to probe a meaningful range of parameter space, while shallow searches to depths of $\approx 20-21$~AB~mag may only be marginally effective. This highlights the key role of $\gtrsim 4-8$-meter aperture telescopes in performing meaningful searches for electromagnetic counterparts to gravitational waves.

\section*{Acknowledgements}

Support for this work was provided by NASA through Einstein Postdoctoral Fellowship grant number PF4-150121. RM acknowledges generous support from the James Arthur Fellowship at NYU. RC acknowledges support from NASA Swift grant NNX16AB04G. EB acknowledges support from NSF grant AST-1411763 and NASA ADA grant NNX15AE50G. BJS is supported by NASA through Hubble Fellowship grant HST-HF-51348.001 awarded by the Space Telescope Science Institute, which is operated by the Association of Universities for Research in Astronomy, Inc., for NASA, under contract NAS 5-26555. This paper includes data gathered with the 6.5 meter Magellan Telescopes located at Las Campanas Observatory, Chile. Based on observations obtained at the Gemini Observatory acquired through the Gemini Observatory Archive and processed using the Gemini IRAF package which is operated by the Association of Universities for Research in Astronomy, Inc., under a cooperative agreement with the NSF on behalf of the Gemini partnership: the National Science Foundation (United States), the National Research Council (Canada), CONICYT (Chile), Ministerio de Ciencia, Tecnología e Innovación Productiva (Argentina), and Ministério da Ciência, Tecnologia e Inovação (Brazil). The United Kingdom Infrared Telescope (UKIRT) is supported by NASA and operated under an agreement among the University of Hawaii, the University of Arizona, and Lockheed Martin Advanced Technology Center; operations are enabled through the cooperation of the East Asian Observatory. We thank the Cambridge Astronomical Survey Unit (CASU) for processing the WFCAM data, and the WFCAM Science Archive (WSA) for making the data available.  VLA observations were obtained under Program 14A-344. The National Radio Astronomy Observatory is a facility of the National Science Foundation operated under cooperative agreement by Associated Universities, Inc. Based on observations made with the NASA/ESA Hubble Space Telescope, obtained from the Data Archive at the Space Telescope Science Institute, which is operated by the Association of Universities for Research in Astronomy, Inc., under NASA contract NAS 5-26555. These observations are associated with program \#13830. This research has made use of the NASA/IPAC Extragalactic Database (NED) which is operated by the Jet Propulsion Laboratory, California Institute of Technology, under contract with the National Aeronautics and Space Administration. This work made use of data supplied by the UK Swift Science Data Centre at the University of Leicester. The scientific results reported in this article are based in part on observations made by the Chandra X-ray Observatory (ObsID: 17594) and data obtained from the Chandra Data Archive (ObsID: 17586). Based on observations collected at the European Organisation for Astronomical Research in the Southern Hemisphere.

{\it Facilities:} Swift (BAT, XRT), Chandra (ACIS-S), Magellan (IMACS), Gemini-South (GMOS), UKIRT (WFCAM), HST, VLA, VLT (HAWK-I)

\end{document}